\documentclass[preprint,12pt]{elsarticle}
\usepackage{booktabs,longtable,ragged2e,array}
\newcolumntype{P}[1]{>{\RaggedRight\arraybackslash}p{#1}}
\usepackage{tabularx}
\newcolumntype{Y}{>{\RaggedRight\arraybackslash}X}
\usepackage{times}
\usepackage{xcolor}
\usepackage{epsfig}
\usepackage{graphicx}
\usepackage{amssymb}
\usepackage{tabularx}
\usepackage{longtable}
\usepackage{longtable, geometry, array}
\usepackage{multicol,multirow}
\usepackage{booktabs}
\usepackage{longtable}
\usepackage{bigstrut}
\usepackage{floatrow}
\usepackage{rotating}
\usepackage{adjustbox}
\usepackage{amsmath}
\usepackage{longtable, booktabs}
\journal{ArXiv}
\begin{document}

\begin{frontmatter}

\title{ The Role of AI in Early Detection of Life-Threatening Diseases: A Retinal Imaging Perspective}

\author[inst1]{Tariq M Khan}

\affiliation[inst1]{organization={Department of Cybersecurity and Digital Forensics},%Department and Organization
            addressline={Center of Excellence in Cybercrime and Digital Forensics, Naif Arab University for Security Sciences },
            city={Riyadh},
            postcode={11452},
            country={Kingdom of Saudi Arabia}}

\author[inst2]{Toufique Ahmed Soomro}
\affiliation[inst2]{organization={Artifical Intelligence and Cyber Future Institute},%Department and Organization
            addressline={Charles Sturt University},
            city={Mitchell, Bathurst},
            postcode={2795},
            state={NSW},
            country={Asutralia}}
            
\author[inst3]{Imran Razzak}
\affiliation[inst3]{organization={Mohamed bin Zayed University of Artificial Intelligence},%Department and Organization
            city={Abu Dhabi},
            country={United Arab Emirates}}

\begin{abstract}
Retinal imaging has emerged as a powerful, non-invasive modality for detecting and quantifying biomarkers of systemic diseases—ranging from diabetes and hypertension to Alzheimer’s disease and cardiovascular disorders—but current insights remain dispersed across platforms and specialties. Recent technological advances in optical coherence tomography (OCT/OCTA) and adaptive optics (AO) now deliver ultra-high-resolution scans (down to 5 µm) with superior contrast and spatial integration, allowing early identification of microvascular abnormalities and neurodegenerative changes. At the same time, AI-driven and machine learning (ML) algorithms have revolutionized the analysis of large-scale retinal datasets, increasing sensitivity and specificity: for example, deep learning models achieve $>$ 90 \% sensitivity for diabetic retinopathy and AUC = 0.89 for the prediction of cardiovascular risk from fundus photographs. The proliferation of mobile health technologies and telemedicine platforms further extends access, reduces costs, and facilitates community-based screening and longitudinal monitoring. Despite these breakthroughs, translation into routine practice is hindered by heterogeneous imaging protocols, limited external validation of AI models, and integration challenges within clinical workflows. In this review, we systematically synthesize the latest OCT / OCT and AO developments, AI / ML approaches, and mHealth / Tele-ophthalmology initiatives, and quantify their diagnostic performance across disease domains. Finally, we propose a roadmap for multicenter protocol standardization, prospective validation trials, and seamless incorporation of retinal screening into primary and specialty care pathways –paving the way for precision prevention, early intervention, and ongoing treatment of life-threatening systemic diseases.
\end{abstract}

\end{frontmatter}

\section{Introduction}
\label{sec:Intro}

\begin{figure}
  \centering
  \includegraphics[width=\textwidth]{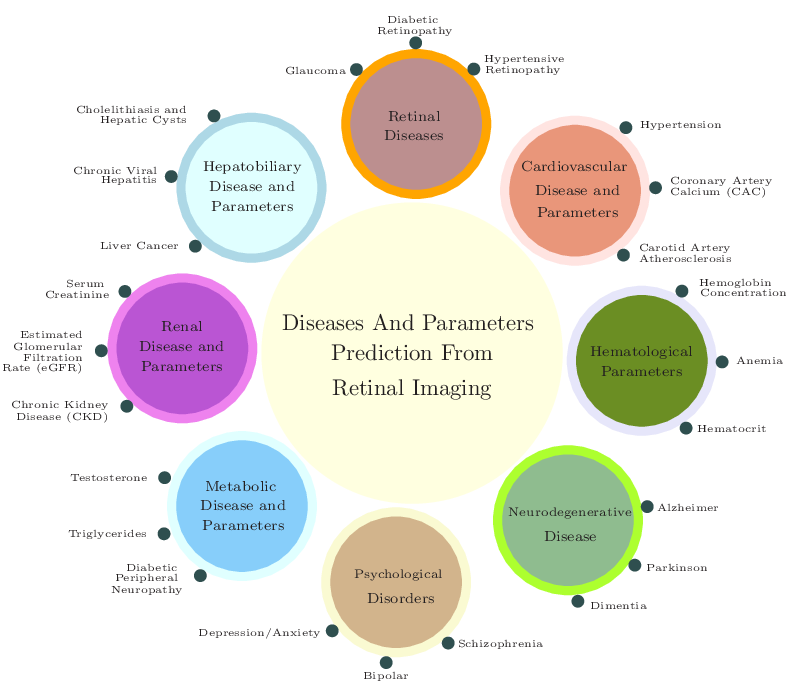} 
  \caption{Bird's-eye view of diseases and parameter prediction from retinal imaging.}
  \label{Birdeye}
\end{figure}

%===== Background: Retina as a window to systemic health =====
The retina serves as a biological indicator of systemic health due to its embryonic, functional, and structural similarities with the main organs such as the heart, brain, and kidneys \cite{WONG2014}. Its microvascular network provides a direct view of circulatory health, allowing in vivo evaluation of systemic changes associated with neurodegenerative, metabolic, and cardiovascular diseases \cite{Flammer2013}. As a non-invasive diagnostic tool, retinal imaging enables the assessment of vascular integrity, neural function, and early pathological alterations in real time \cite{WONG2014,Patton2006}. In many chronic diseases, retinal abnormalities can appear before clinical symptoms manifest, making them a valuable tool for early diagnosis and disease monitoring \cite{imtiaz2021screening,khan2022width,khan2022t}.

%===== Imaging modalities & AI integration =====
Retinal imaging plays an important role in the analysis of both the structural and functional aspects of the retina, offering critical biomarkers for the diagnosis and tracking of systemic diseases, including diabetes, Alzheimer’s disease, hypertension, and coronary artery and cerebrovascular diseases (CACVD) \cite{soomro2018impact,rehman2019multi,tabassum2020cded,khan2020residual,iqbal2022recent}. The evolution of advanced imaging technologies—particularly adaptive optics (AO) and optical coherence tomography (OCT)—has significantly improved high-resolution retinal visualization, allowing a better understanding of disease progression and risk assessment. The integration of machine learning (ML) and artificial intelligence (AI) further enhances diagnostic precision by detecting systemic biomarkers beyond predefined retinal features \cite{Wong2001,Ting2019,Mitani2020}. Deep learning models have demonstrated the ability to predict cardiovascular risk factors and detect systemic conditions directly from retinal images \cite{Kim2020,Chatterjee2002,Zhang2023}.

%===== Retinal vascular architecture =====
The retinal vasculature is precisely organized to regulate blood flow, and deviations from its normal architecture often signal underlying disease \cite{soomro2017computerised,rehman2019multi,khawaja2019improved,khawaja2019multi}. Features such as vessel tortuosity, aberrant bifurcation patterns, and altered arteriovenous ratios have been linked to conditions such as hypertension and other cardiovascular disorders \cite{Poplin2018,Zamir1976}. A comprehensive list of these vascular changes is shown in Figure~\ref{Birdeye}. Studies suggest that mutations in COL4A1, a gene involved in collagen synthesis, may contribute to microvascular abnormalities and the risk of hemorrhagic stroke \cite{Gould2006}. Fractal analysis provides a quantitative measure of vascular complexity, providing a deeper understanding of both retinal development and microvascular alterations \cite{Liew2008,Mainster1990}.

%===== Microcirculation and blood‐flow assessment =====
The evaluation of retinal microcirculatory blood flow is crucial, as it provides valuable information on cardiovascular risk factors and systemic vascular health. Techniques such as scanning laser Doppler flowmetry and scanning laser ophthalmoscopy allow real-time quantification of microcirculatory dynamics - capabilities unmatched by other methods \cite{Harazny2007,Michelson2007}. When combined with vessel imaging, these approaches facilitate precise measurements of arteriolar and venular wall thickness, establishing a robust framework for the assessment of microvascular health. Retinal vasoreactivity studies have also linked vascular dysfunction to hypertension and diabetes mellitus, strengthening the role of the retina in monitoring systemic diseases \cite{Riva2005}.

%===== Neurodegenerative biomarkers in the retina =====
Investigations of neurodegenerative disorders highlight retinal imaging as a potential biomarker of conditions such as Alzheimer’s disease (AD). Retinal abnormalities, including amyloid plaques and neurofibrillary tangles, have been explored as non-invasive indicators of early neurodegeneration, although more studies are required to validate these correlations \cite{Wagner2020,Barriada2022,Syed2025}.  

%===== Genetic and macro vs. microvascular gaps =====
Similarly, genetic research has sought to clarify the molecular basis of vascular diseases and their association with cardiovascular risk. However, most studies have focused on macrovascular conditions, such as atherosclerosis, with limited attention to microvascular genetic determinants, a gap that calls for a deeper exploration of hereditary influences on vascular integrity \cite{Xing2006,Witt2006}. No prior review has systematically unified AI-enhanced retinal oculomics across both neurological and cardiovascular disease domains to provide an integrated framework for clinical translation.

%===== Aim and scope =====
This review provides a comprehensive assessment of the role of retinal imaging in detecting and monitoring systemic diseases, with a focus on cardiovascular, neurodegenerative, metabolic, and hematological disorders. It analyzes retinal biomarkers and their potential as non-invasive indicators of disease progression, offering a detailed evaluation of how retinal imaging can aid in early diagnosis. The paper highlights advances in AI and deep learning (DL), showcasing their ability to enhance diagnostic accuracy by identifying systemic biomarkers beyond predefined retinal features. It emphasizes the role of multimodal imaging techniques, including OCT, OCT angiography (OCTA), and fundus photography, in improving disease detection and monitoring. The study also examines challenges in integrating retinal imaging into routine clinical practice, such as the need for standardized imaging protocols, external validation of AI models, and improved clinical adoption. In addition, it explores the potential of retinal imaging in precision medicine, telemedicine, and large-scale disease detection, emphasizing its transformative impact on non-invasive diagnostics.

%===== Manuscript organization =====
The remainder of the paper is organized as follows. Section~1 introduces the concept of retinal imaging as a noninvasive diagnostic tool, highlighting recent advancements in AI and DL that have improved disease prediction. Section~2 focuses on common retinal diseases, such as hypertensive retinopathy, retinal vein occlusion (RVO), glaucoma, and diabetic retinopathy (DR), discussing their association with broader systemic health conditions. Section~3 explores retinal biomarkers and OCT in the identification of neurodegenerative diseases (AD, PD, MS), highlighting the potential for early detection. Section~4 examines the link between retinal vascular changes and cardiovascular diseases (CVD) -hypertension, atherosclerosis, stroke risk - and the role of AI-driven imaging in the assessment of CVD risk. Section~5 expands to other systemic diseases, including hematologic, metabolic, renal, and hepatobiliary disorders, evaluating retinal imaging as a reliable biomarker. Section~6 looks at future potential—integrating AI, multimodal imaging, and clinical applications while addressing challenges (standardization, validation, and clinical implementation). Finally, Section~7 summarizes key findings and underscores the need for further research and standardization to establish retinal imaging as a valuable clinical tool.

\section{Common Retinal Diseases} \label{sec:RD}
Retinal vascular diseases are among the most prevalent eye conditions, often associated with systemic disorders such as hypertension, diabetes, cardiovascular disease (CVD), and metabolic dysfunctions. These diseases are primarily the result of vascular abnormalities, chronic inflammation, and degenerative changes that alter blood flow and retinal function. Common retinal conditions include hypertensive retinopathy, retinal vein occlusion (RVO), central retinal artery occlusion (CRAO), and diabetic retinopathy (DR), all of which can lead to gradual vision loss if not identified and managed promptly. Factors such as aging, genetic predisposition, metabolic imbalances, and vascular complications significantly contribute to their development and progression.

\begin{table}[h]
  \centering
  \caption{Imaging Modalities, Key Retinal Biomarkers, and Associated Systemic Risks}
  \begin{tabular}{p{3cm}p{3.5cm}p{4cm}p{3cm}}
    \toprule
    \textbf{Disease} & \textbf{Imaging Modality} & \textbf{Key Retinal Biomarkers} & \textbf{Systemic Link} \\
    \midrule
    Hypertensive Retinopathy & Fundus, OCT/OCTA & Arteriolar narrowing, AV nicking, vessel tortuosity & Hypertension, LV hypertrophy, stroke risk \\
    Proliferative Diabetic Retinopathy & Fundus, OCT & Neovascularization, macular edema, capillary dropout & Diabetes complications, nephropathy \\
    Central Retinal Artery Occlusion & Fundus, OCT & Cherry‐red spot, inner retinal swelling & Stroke risk, carotid stenosis \\
    Retinal Vein Occlusion & Fundus, OCTA & Venular dilation, hemorrhages, nonperfusion & Venous thromboembolism, CVD mortality \\
    \bottomrule
  \end{tabular}
  \label{tab:retinal-disease-summary}
\end{table}

With the advancement of retinal imaging technologies, including optical coherence tomography (OCT), OCT angiography (OCTA), and fundus photography, the ability to detect and monitor retinal abnormalities has greatly improved, allowing early diagnosis and targeted interventions. In addition, the incorporation of artificial intelligence (AI) and deep learning (DL) has improved automated disease screening, risk assessment, and diagnostic accuracy. Since retinal health is closely related to systemic conditions, utilizing retinal imaging for disease detection can play a crucial role in both ophthalmic and general healthcare. The following sections provide a detailed discussion of different types of common retinal diseases, their underlying mechanisms, and their clinical implications.
\begin{figure}[htbp]
  \centering
  \includegraphics[width=0.99\textwidth]{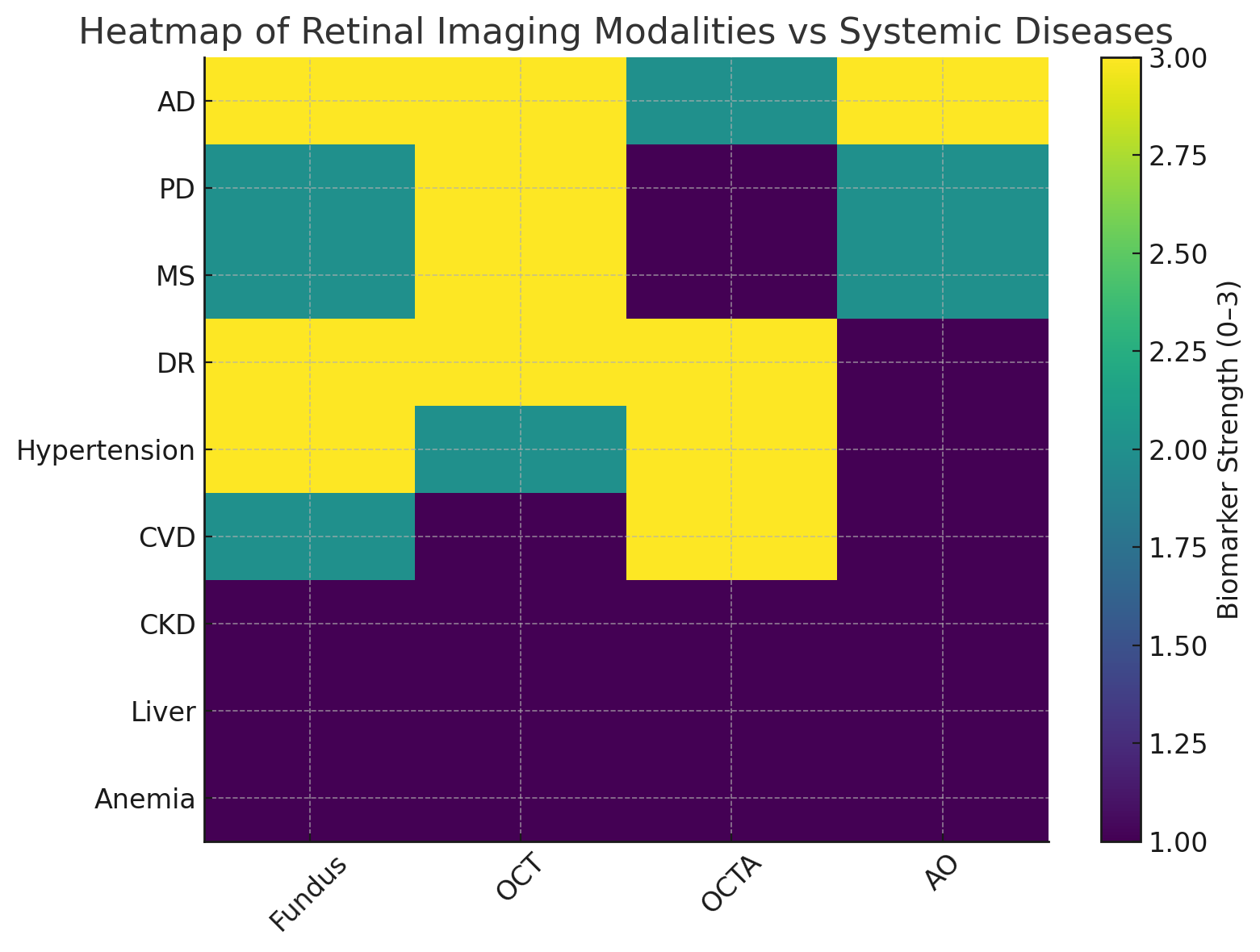}
  \caption{Heatmap of retinal imaging modalities (fundus, OCT, OCTA, AO) versus systemic disease categories (AD, PD, MS, DR, hypertension, CVD, CKD, hepatobiliary disease, anemia). Colors indicate relative biomarker strength from none (0) to strong (3).}
  \label{fig:modality_disease_heatmap}
\end{figure}

As shown in Figure~\ref{fig:modality_disease_heatmap}, color fundus photography and OCT offer the strongest biomarkers for diabetic retinopathy and hypertension, while OCTA provides superior perfusion metrics for cardiovascular disease. Adaptive optics delivers high‐resolution imaging of neurodegenerative signatures, particularly in Alzheimer's disease, highlighting its emerging role in early diagnosis. These strengths of modality-disease are drawn from key quantitative studies (e.g., Poplin et al. \cite{Poplin2018}, Ma \emph{et al.} \cite{Ma2023}, Sabanayagam \emph{et al.} \cite{Sabanayagam2020}).

\subsection{Hypertensive Retinopathy}
\begin{figure}[h!]
  \centering
  \includegraphics[width=1\textwidth]{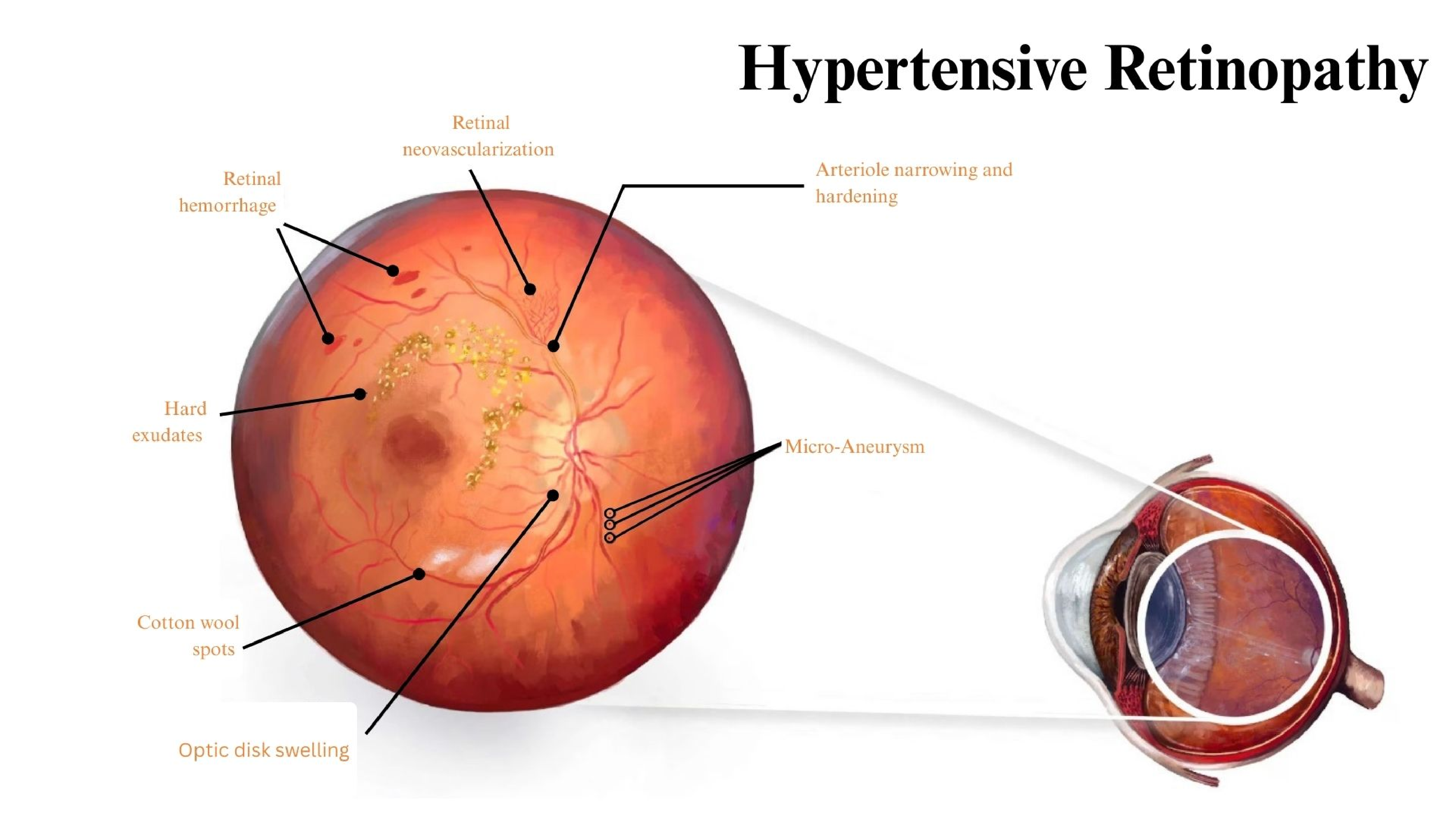}
  \caption{Representative retinal changes observed in hypertensive retinopathy. Key features include retinal hemorrhages, neovascularization, arteriolar narrowing and bundling, micro-aneurysms, hard exudates, cotton-wool spots, and optic disc swelling. These vascular abnormalities reflect the impact of systemic hypertension on the retinal microvasculature and are critical indicators for the diagnosis and monitoring of hypertensive damage.}\label{Hper}
\end{figure}

Hypertensive retinopathy (HR) is a vascular condition resulting from chronic hypertension, leading to progressive damage to the retinal blood vessels \cite{soomro2016automatic, khan2016automatic}, as shown in Figure \ref{Hper}. Prolonged high blood pressure exerts an excessive force on the retinal microcirculation, causing arteriolar constriction, thickening of the vascular wall, and increased rigidity, ultimately altering blood flow and oxygen delivery to the retina. As the condition worsens, structural changes in the vasculature can cause retinal hemorrhages, lipid exudates, macular edema, and optic nerve swelling, significantly affecting vision \cite{Wong2004,wong2007eye}. The severity of HR is often proportional to the duration and intensity of hypertension, making retinal evaluation a valuable indicator of systemic vascular health.

The progression of HR follows a sequence of vascular remodeling, beginning with arteriolar narrowing due to persistent vasospasm. Over time, the arterial walls undergo thickening and hyaline sclerosis, contributing to arteriovenous (AV) nicking, where stiffened arteries compress adjacent veins. This compression can result in venous congestion, ischemia, and possible retinal infarction, further exacerbating tissue damage. In advanced stages, flame-shaped hemorrhages, cotton wool spots, and hard exudates appear due to capillary leakage and infarction of the nerve fiber layer, indicating severe end-organ damage \cite{Chatterjee2002}. These retinal abnormalities often parallel hypertensive complications in other organs, reinforcing the importance of early detection and management.

Several classification systems have been developed to assess the severity of HR. The Keith-Wagener-Barker classification system categorizes the disease into four stages, ranging from mild arteriolar narrowing (grade 1) to severe vascular damage with swelling of the optic disc (grade 4). Early HR is often asymptomatic and detectable only through fundoscopic examination, while more advanced stages can lead to blurred vision, loss of the visual field, and ischemic retinal changes. Furthermore, Grosso \emph{et al.} refined the HR classification focusing on vascular stiffness, AV compression, and structural changes, correlating these characteristics with systemic blood pressure levels \cite{Grosso2005}.

The clinical importance of HR extends beyond ocular manifestations, as retinal vascular abnormalities reflect systemic hypertension-related damage in other organs. Moderate to severe HR has been associated with an increased risk of stroke, myocardial infarction, and chronic kidney disease (CKD), highlighting the importance of retinal examination in the treatment of hypertensive patients. Advances in OCT and OCTA have improved early detection of microvascular damage, allowing better risk assessment and timely intervention.

Effective HR management requires early identification and appropriate treatment of hypertension to prevent irreversible retinal damage. Regular blood pressure control, lifestyle modifications, and antihypertensive therapy are key to reducing disease progression and preventing vision loss. Given the widespread impact of hypertension, integrating retinal imaging into routine healthcare practices provides a non-invasive and reliable method for assessing vascular health and systemic disease risk.

\subsection{Retinal Vein Occlusion (RVO)}
\begin{figure}[h!]
  \centering
  \includegraphics[width=1\textwidth]{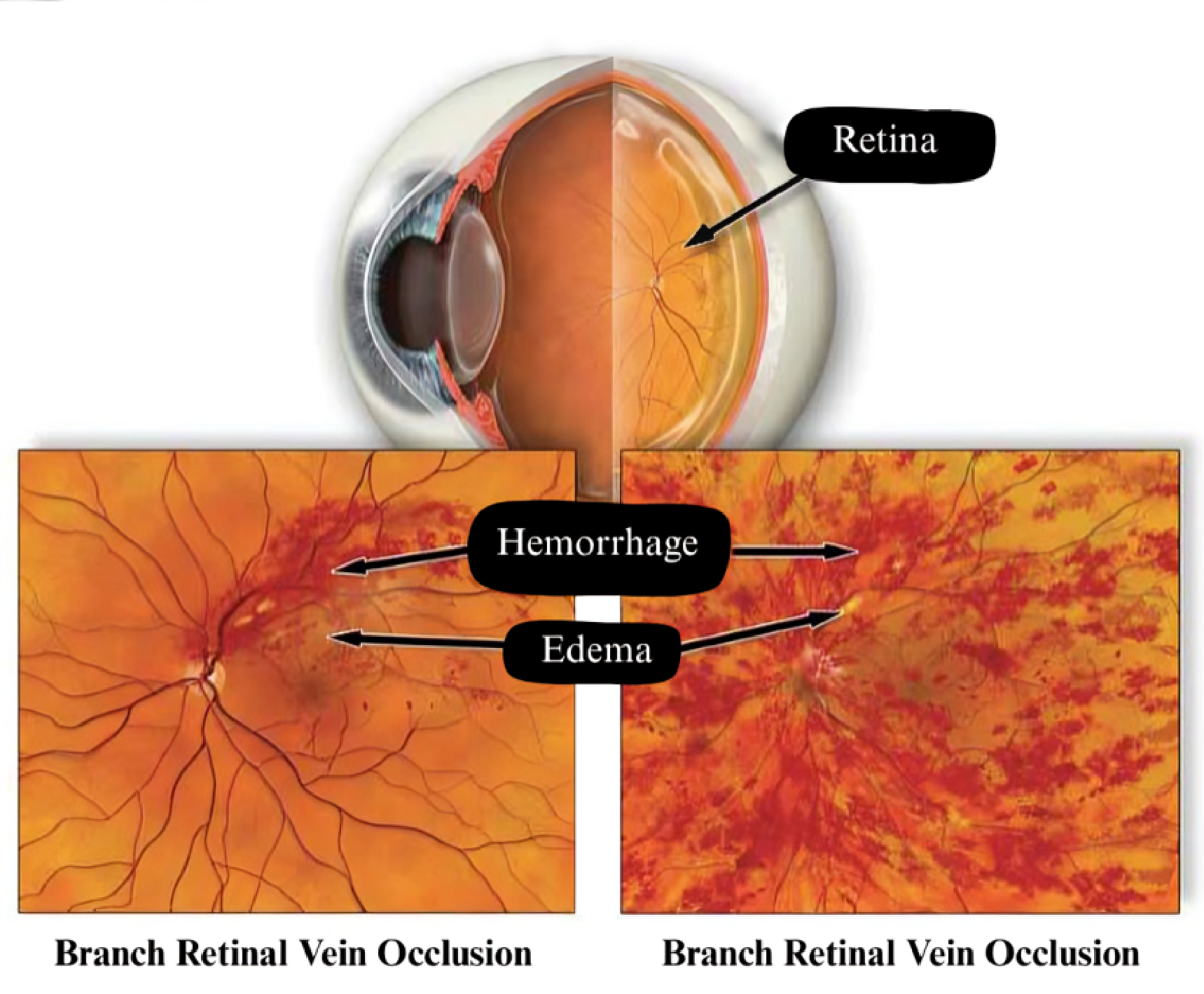}
  \caption{Representative retinal images showing Branch Retinal Vein Occlusion (BRVO). Key pathological features include hemorrhage and edema within the retinal layers, resulting from obstruction of the retinal venous outflow. These changes can lead to vision impairment and are critical indicators in the diagnosis and management of BRVO.}\label{rvo}
\end{figure}

RVO is a vascular disorder that occurs due to obstruction of the retinal veins, leading to impaired venous drainage, increased intravascular pressure, and retinal ischemia. It is a leading cause of vision loss associated with systemic conditions such as hypertension, diabetes mellitus (DM), dyslipidemia as shown in Figure \ref{rvo}, and systemic vasculopathy \cite{Marcucci2011}. These underlying conditions contribute to vascular dysfunction, increasing the likelihood of thrombotic events and blood flow obstruction in the retinal venous system. The occlusion can manifest in two primary forms: branch retinal vein occlusion (BRVO) and central retinal vein occlusion (CRVO), classified according to the location of the blockage.

The pathophysiology of RVO involves the formation of thrombuses within the retinal venous circulation, leading to stasis, hemorrhage, and edema. The obstruction results in increased venous pressure, which disrupts normal blood flow and causes capillary non-perfusion, vessel leakage, and macular edema. Over time, this condition can cause retinal ischemia and neovascularization, which, if left untreated, can cause severe visual impairment. BRVO occurs when a branch of the central retinal vein becomes blocked, usually at an arteriovenous crossing site, leading to localized retinal hemorrhages and ischemia. In contrast, CRVO involves complete blockage of the central retinal vein, resulting in widespread retinal hemorrhages, optic disc swelling, and extensive macular edema, often with a poorer prognosis due to its greater impact on retinal circulation.

Several risk factors contribute to RVO, including systemic hypertension, atherosclerosis, hypercoagulability, and endothelial dysfunction. The association between RVO and cardiovascular disease (CVD) highlights the need for a comprehensive assessment of vascular health in affected individuals. Hyperviscosity syndromes, including polycythemia and thrombophilic disorders, further increase the risk of thrombotic events within the retinal circulation. Furthermore, glaucoma and optic disc abnormalities can increase susceptibility to RVO by contributing to vascular compression and impaired venous outflow \cite{Lee2010}. 

The clinical presentation of RVO varies based on the extent and severity of venous occlusion. Patients may experience sudden and painless vision loss, typically in one eye, accompanied by blurred vision, distorted central vision, and floating vision. Fundoscopic examination reveals retinal hemorrhages, tortuous dilated veins, cotton-wool spots, and optic disc edema, depending on the severity of the condition. Fluorescein angiography (FA) and OCT play a crucial role in the diagnosis and monitoring of RVO, allowing the assessment of vascular perfusion, macular edema, and ischemic changes. Advances in OCTA have further improved the ability to detect capillary non-perfusion and subtle vascular abnormalities, aiding in early diagnosis and intervention.

Regular imaging and angiographic visualization are essential for tracking disease progression and response to treatment. Before complete venous occlusion occurs, certain vascular alterations can be observed, such as changes in venous caliber, collateral vessel formation, and abnormalities of the arteriovenous crossing \cite{Lee2013}. These subtle alterations serve as early indicators of disease progression, highlighting the importance of routine retinal evaluations in at-risk individuals.

The management of RVO depends on the severity of the condition and the presence of complications such as macular edema and neovascularization. Intravitreal anti-vascular endothelial growth factor (anti-VEGF) therapy, corticosteroids, and laser photocoagulation remain the primary treatment options for reducing macular edema and preventing further visual deterioration. Controlling systemic risk factors, including regulation of blood pressure, glycemic management, and lipid control, is equally essential to prevent recurrence and minimize long-term ocular and systemic complications. Given the strong association between RVO and systemic vascular disorders, patients diagnosed with RVO should undergo comprehensive cardiovascular and hematological evaluations to assess their overall risk of thrombotic and systemic health.

\subsection{Glaucoma}
\begin{figure}[h!]
  \centering
  \includegraphics[width=1\textwidth]{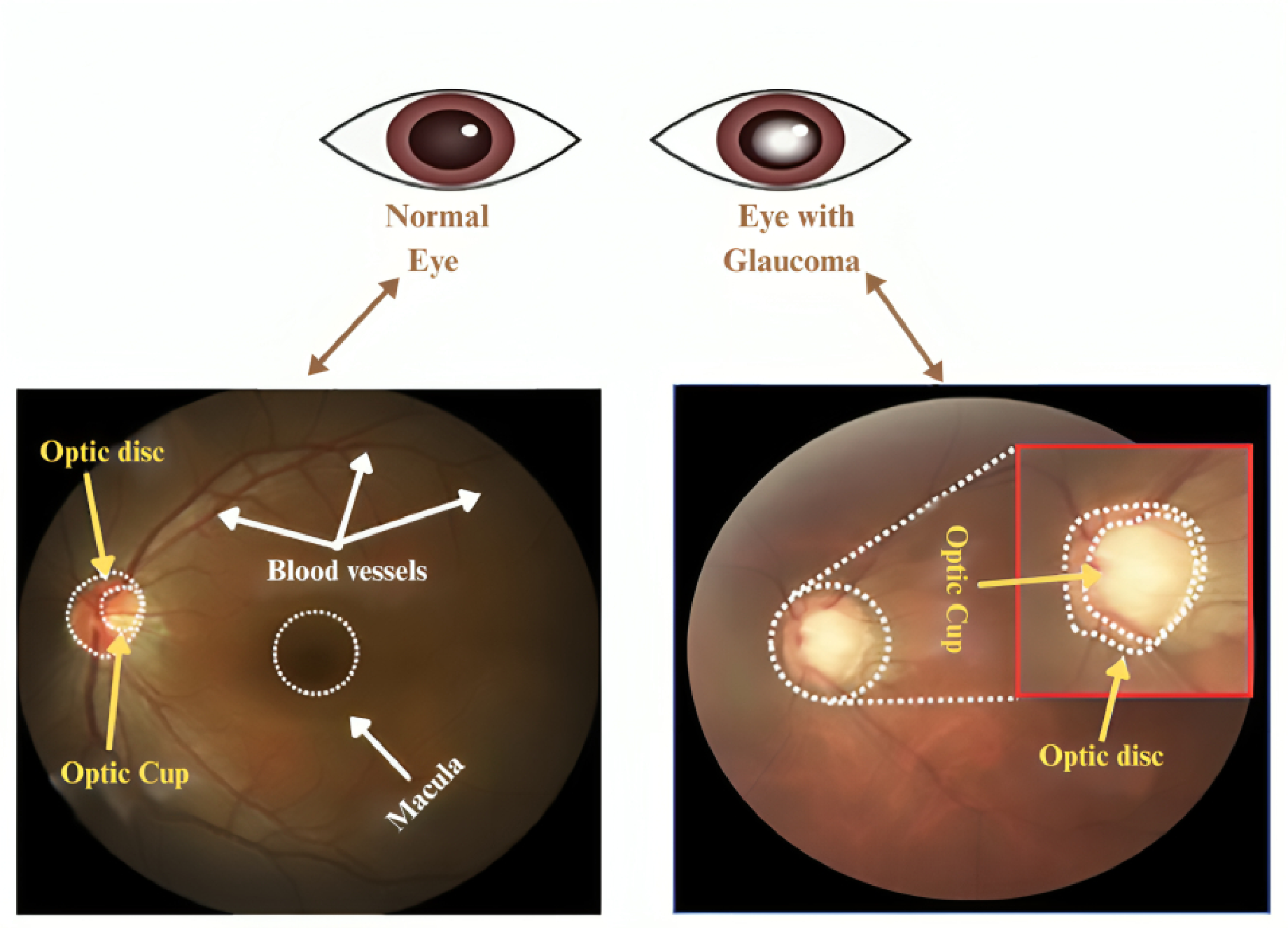}
  \caption{Comparison of a normal eye and an eye affected by glaucoma, highlighting structural differences in the optic nerve region. In the glaucomatous eye, the optic cup appears enlarged relative to the optic disc, indicating optic nerve head cupping \. This structural change is a hallmark of glaucoma, often resulting from increased intraocular pressure and associated with progressive retinal ganglion cell loss.}\label{Glauocma}
\end{figure}

Glaucoma is one of the leading causes of irreversible blindness, particularly in developing countries, and is characterized primarily by progressive degeneration of the optic nerve and retinal ganglion cells \cite{Quigley2006,Abdullah2021,Imtiaz2021}. The condition often develops silently, with vision loss occurring gradually and becoming noticeable only in advanced stages. A key pathological characteristic of glaucoma is optic disc cupping, which results from the loss of retinal nerve fibers and leads to permanent visual field defects, as shown in Figure \ref{Glauocma}. If not treated, glaucoma-induced damage can progress to complete blindness.

Multiple risk factors contribute to the development and progression of glaucoma, including elevated intraocular pressure (IOP), systemic hypertension, obesity, migraine headaches, genetic predisposition, and vascular dysregulation. Although increased IOP is the most recognized risk factor, some forms of glaucoma occur even with normal IOP, suggesting the involvement of vascular and neurodegenerative mechanisms. The disease is classified into three primary types: open-angle glaucoma (OAG), closed-angle glaucoma (CAG), and normal-tension glaucoma (NTG).

OAG is the most prevalent form, often asymptomatic in the early stages. It results from a gradual obstruction in the trabecular meshwork, which leads to inadequate drainage of the aqueous humor and progressive optic nerve damage. Due to its slow progression, patients may not experience noticeable vision loss until substantial damage has occurred. Without intervention, OAG can result in peripheral vision loss that progresses to central vision impairment.

CAG, on the other hand, occurs when the drainage angle between the iris and cornea becomes blocked, leading to a sudden increase in IOP. This form of glaucoma often presents with acute symptoms, including severe eye pain, headaches, nausea, and blurred vision. If not treated promptly, CAG can cause rapid optic nerve damage and vision loss. Unlike OAG, CAG requires urgent medical intervention to prevent permanent visual impairment.

NTG is unique in that it occurs without elevated IOP, suggesting that impaired blood flow to the retinal and optic nerve plays a crucial role in the progression of the disease. Patients with NTG often exhibit vascular dysregulation, leading to reduced optic nerve perfusion. This form of glaucoma is commonly associated with systemic conditions such as low blood pressure, nocturnal hypotension, and vasospastic disorders. Despite normal IOP levels, optic nerve damage and loss of visual field progress similarly to other types of glaucoma.

The diagnosis of glaucoma is based on multiple clinical parameters, including IOP measurements, retinal nerve fiber layer thickness (RNFL) analysis, optic nerve head imaging, and visual field testing \cite{Thomas2006} . OCT and OCTA provide high-resolution imaging of the optic nerve head and retinal microvasculature, allowing early detection of glaucomatous changes. Additional evaluations, such as cup-to-disc ratio analysis, anterior chamber angle evaluation, and corneal thickness measurement, further aid in diagnosing and monitoring disease progression.

Early detection and management of glaucoma are critical to preventing irreversible vision loss. Current treatment strategies focus on lowering IOP through medications, laser therapy, or surgical interventions. Topical eye drops, including prostaglandin analogs, beta-blockers, and carbonic anhydrase inhibitors, are commonly used to reduce the production of aqueous humor or enhance the flow of the odor. In cases where medications are ineffective, laser trabeculoplasty or surgical procedures, such as trabeculectomy or minimally invasive glaucoma surgery (MIGS), may be required to preserve optic nerve function.

Since glaucoma is a lifelong disease, regular screening and early intervention play a vital role in reducing the burden of the disease. With the advancement of AI-based retinal imaging, automated screening tools can improve early detection, allowing timely treatment and better visual outcomes. Given the silent progression of OAG and NTG, integrating glaucoma screening into routine ophthalmic care is essential to prevent preventable blindness and preserve vision quality.

\subsection{Central Retinal Artery Occlusion (CRAO)}
\begin{figure}[h!]
  \centering
  \includegraphics[width=1\textwidth]{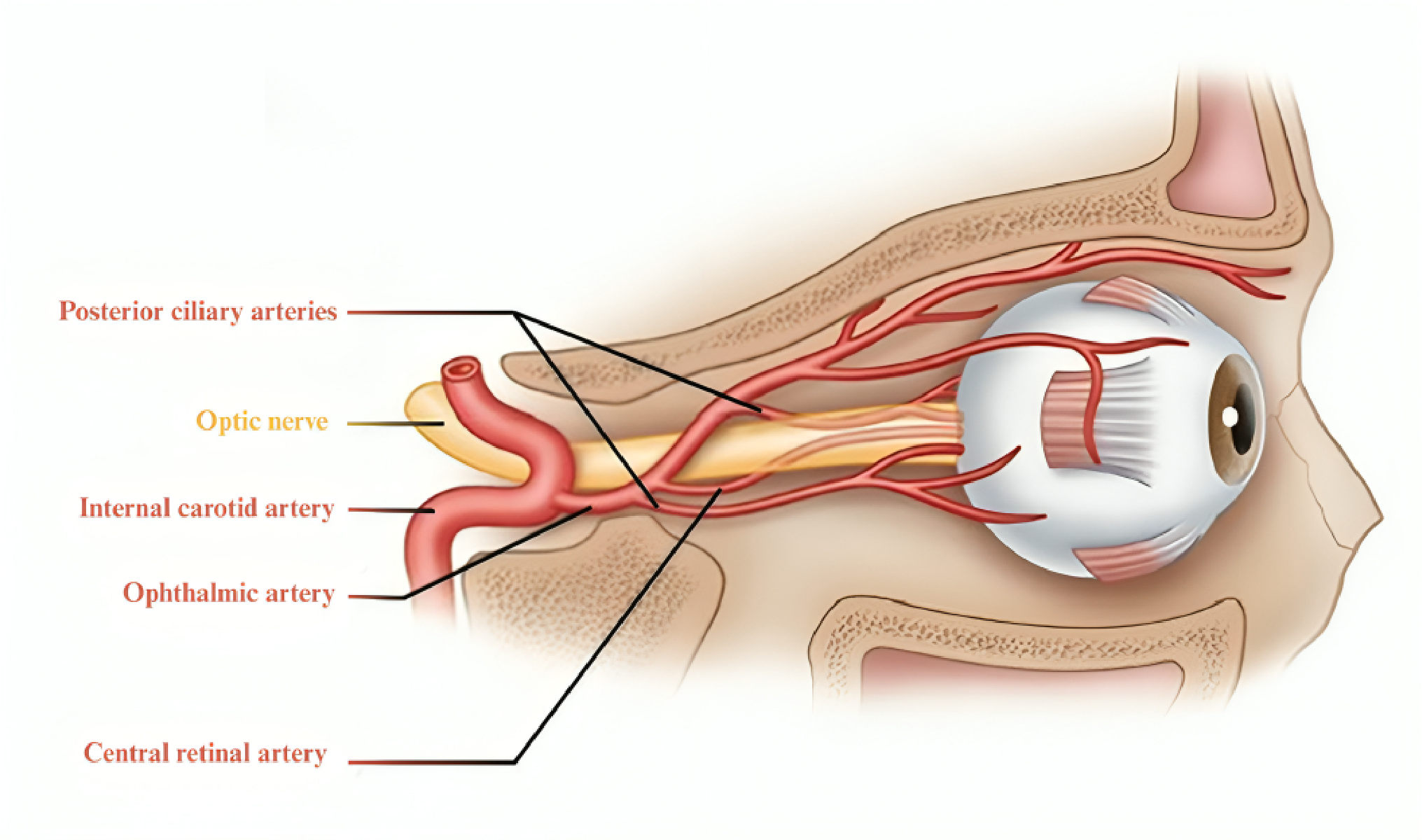}
  \caption{Anatomical pathway of the ocular blood supply and optic nerve. The image depicts the key vascular structures including the internal carotid artery, ophthalmic artery, central retinal artery, and posterior ciliary arteries, along with the optic nerve. These vessels play a vital role in maintaining retinal and optic nerve function, and their compromise can lead to serious ocular and neurodegenerative conditions.}\label{CRAO}
\end{figure}

CRAO is an ocular emergency characterized by sudden and painless vision loss due to an acute blockage of the central retinal artery, which supplies oxygenated blood to the inner layers of the retina, as shown in Figure \ref{CRAO}. This condition is considered the ocular equivalent of a stroke and requires immediate medical intervention to minimize permanent retinal ischemia and vision loss. The most common causes of CRAO include embolic events in the internal carotid artery, cardiac embolism, or thrombosis, leading to retinal vascular occlusion and reduced blood supply to the eye. Atherosclerosis, a systemic disease characterized by the accumulation of fatty plaque in the arteries, is a major risk factor for CRAO and is often associated with hypertension, DM, and hyperlipidemia \cite{Varma2013}.

The pathophysiology of CRAO involves an interruption in retinal blood flow, resulting in acute retinal ischemia and tissue hypoxia. Since the retina has one of the highest metabolic demands in the body, even a brief period of ischemia can cause irreversible neuronal damage. Within minutes to hours after arterial occlusion, the retina develops a pale and edematous appearance, except for the fovea, which remains red due to the underlying choroidal circulation, creating the classic ``cherry-red spot'' at the macula. Over time, chronic ischemia leads to retinal atrophy, pallor of the optic nerve, and permanent visual impairment.

CRAO is most frequently seen in older adults, particularly those with underlying CVD, carotid artery stenosis, atrial fibrillation, or coagulation disorders. The condition is often associated with TIAs, giant cell arteritis (GCA), and hypercoagulable states, which makes it essential for clinicians to perform a thorough systemic evaluation when diagnosing CRAO. The presence of embolic material in the retinal vasculature can indicate an increased risk of stroke and cardiovascular events, which requires urgent neurological and vascular evaluation.

The clinical presentation of CRAO is characterized by sudden, severe, painless monocular vision loss, often described as a ``curtain descending over the eye''. Some patients may experience transient episodes of visual disturbance, known as amaurosis fugax, prior to complete occlusion. Fundoscopic examination typically reveals a pale retina with attenuated retinal vessels and a cherry-red macula, which is considered a diagnostic hallmark of CRAO.

FA plays a crucial role in confirming the diagnosis of CRAO, as it can detect delayed or absent arterial filling in the affected retina. Additional imaging modalities, such as OCT and OCTA, provide further information on retinal ischemia and structural damage, allowing for a more comprehensive assessment of the disease. Given the high risk of cerebrovascular events in patients with CRAO, it is recommended to urgently refer to carotid Doppler ultrasound, echocardiography, and brain imaging (MRI or CT angiography) to identify potential embolic sources.

Treatment of CRAO remains challenging due to the narrow therapeutic window for intervention. Various approaches have been attempted to restore retinal perfusion, including ocular massage, anterior chamber paracentesis, and hyperbaric oxygen therapy (HBOT). In some cases, thrombolytic therapy has been explored as a possible treatment option; however, its use remains controversial due to associated risks and limited efficacy. Management of systemic risk factors, including control of blood pressure, anticoagulation from embolic sources, and lifestyle modifications, is essential to prevent recurrent vascular events.

Since CRAO is associated with a high risk of systemic thromboembolic disease, all patients should undergo a comprehensive cardiovascular evaluation to assess their risk of stroke. Collaboration between ophthalmologists, neurologists, and vascular specialists is essential to ensure optimal patient care and long-term management. Despite advances in diagnostic imaging and acute stroke management, vision loss in CRAO remains largely irreversible, underscoring the need for early detection and reduction in systemic risk to prevent future vascular complications.

\subsection{Diabetic Retinopathy (DR)}
\begin{figure}[h!]
  \centering
  \includegraphics[width=0.95\textwidth]{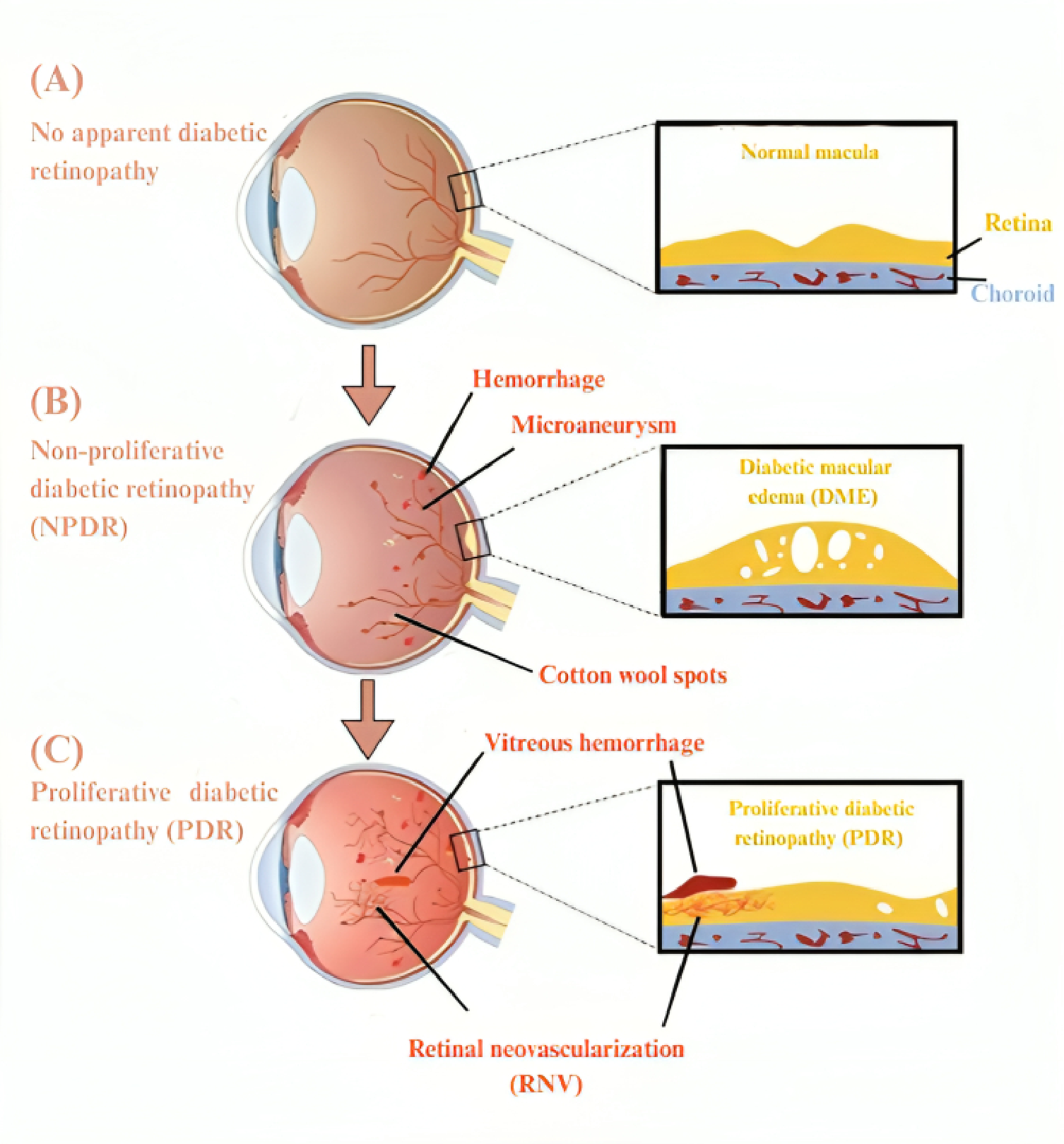}
  \caption{Progression stages of diabetic retinopathy. (A) A healthy retina with no apparent diabetic retinopathy. (B) Non-proliferative diabetic retinopathy (NPDR), characterized by microaneurysms, hemorrhages, cotton wool spots, and diabetic macular edema (DME). (C) Proliferative diabetic retinopathy (PDR), marked by retinal neovascularization (RNV) and vitreous hemorrhage. These progressive changes reflect worsening retinal damage due to prolonged hyperglycemia and are critical for staging and managing diabetic eye disease.}\label{DR}
\end{figure}

Diabetic retinopathy (DR) is a major global health concern and a leading cause of vision impairment and blindness among individuals with diabetes mellitus (DM). Studies indicate that up to 90\% of individuals with DM for more than two decades develop some form of DR \cite{Lee2015}. DM of type 1 and type 2 predispose individuals to this sight-threatening complication, primarily due to chronic hyperglycemia-induced damage to the retinal microvasculature. Uncontrolled blood glucose levels contribute to vascular dysfunction, increased capillary permeability, and pathological neovascularization, which ultimately leads to retinal tissue damage and progressive visual loss as shown in Figure \ref{DR}.

The pathogenesis of DR is driven by metabolic and inflammatory processes that disrupt normal retinal vascular homeostasis. Persistent hyperglycemia promotes the accumulation of advanced glycation end products (AGEs), triggering oxidative stress, chronic inflammation, and endothelial dysfunction. These pathological mechanisms lead to the loss of pericytes, weakening the retinal capillaries and increasing their fragility. As a result, vascular leakage, microaneurysm formation, and ischemic damage occur, progressively impairing retinal function \cite{Duh2017}.

DR progresses through different stages. Non-proliferative diabetic retinopathy (NPDR), the initial phase, is marked by microaneurysms, intraretinal hemorrhages, venous dilation, and capillary dropout. As NPDR progresses, the breakdown of the blood-retinal barrier (BRB) leads to diabetic macular edema (DME), which is the most common cause of DM-related vision loss. In its more severe form, proliferative diabetic retinopathy (PDR), extensive ischemia triggers excessive vascular endothelial growth factor (VEGF) production, causing pathological neovascularization. These fragile vessels proliferate, leading to recurrent hemorrhages, fibrovascular proliferation, and tractional retinal detachment, which can result in irreversible blindness \cite{Claesson2015}.

VEGF plays a pivotal role in the progression of DR by stimulating abnormal angiogenesis and increasing vascular permeability. This results in fluid accumulation in the macula, further contributing to DME, which significantly affects central vision. The central retina, particularly the foveal region, is frequently affected in DME, leading to blurred vision and visual distortion \cite{Claesson2015}.

Early detection and timely intervention are essential in the management of DR. Advanced diagnostic and screening tools, including fundus photography, fluorescein angiography (FA), optical coherence tomography (OCT), and OCT angiography (OCTA), facilitate the assessment of retinal damage and disease severity. Incorporation of AI-driven automated DR screening has further improved early detection and risk stratification, especially in remote and resource-limited areas.

Current therapeutic strategies for DR focus on limiting disease progression and minimizing retinal damage. The most widely used treatments include:

\begin{itemize}
    \item Intravitreal anti-VEGF injections (e.g., bevacizumab, ranibizumab, aflibercept) to suppress abnormal neovascularization and reduce macular edema.
    \item  Corticosteroids are used to counteract inflammatory damage and improve vascular stability.
    \item Laser photocoagulation (panretinal photocoagulation and focal laser therapy) to seal leaky blood vessels and prevent further retinal deterioration.
    \item Vitrectomy surgery for cases with vitreous hemorrhage or tractional retinal detachment \cite{Duh2017}.
\end{itemize}

Long-term management of DR requires strict glycemic control, blood pressure regulation, and lipid management to mitigate systemic risk factors. Regular retinal examinations are essential, as early diagnosis and intervention significantly reduce the likelihood of severe visual impairment. Given the increasing prevalence of DM worldwide, the integration of retinal imaging into routine diabetes care remains a crucial step to preserve vision and prevent blindness.

\section{Neurodegenerative Diseases} \label{sec}

\begin{figure}
  \centering
  \includegraphics[width=1\textwidth]{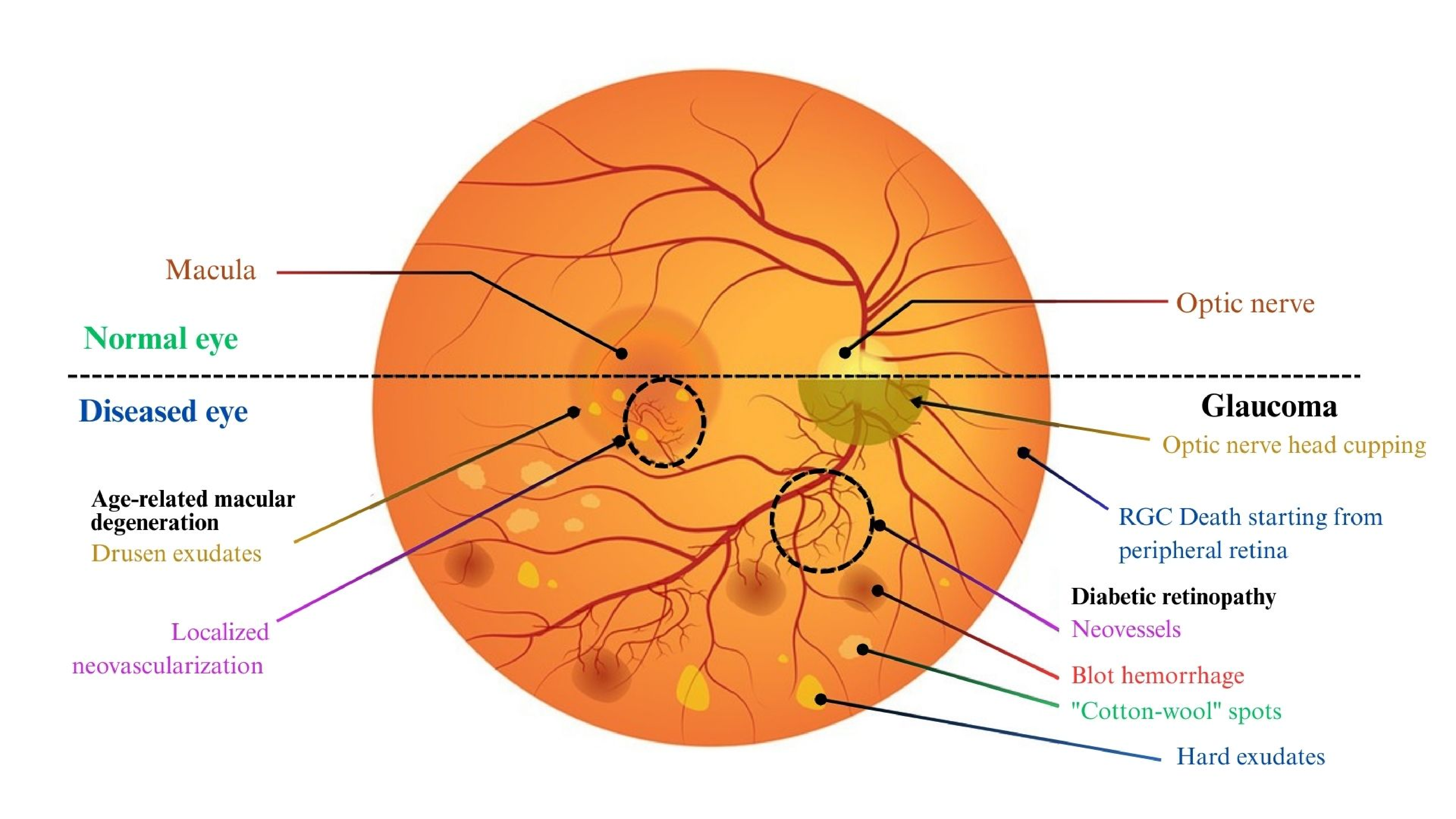}
  \caption{Key retinal features associated with neurodegenerative diseases. The illustration highlights distinctions between normal and diseased eyes, showing abnormalities such as age-related macular degeneration (drusen exudates and localized neovascularization), glaucomatous optic nerve head cupping, retinal ganglion cell (RGC) death starting from the peripheral retina, and diabetic retinopathy markers including neovascularization, blot hemorrhages, "cotton-wool" spots, and hard exudates. These retinal changes serve as critical biomarkers for early detection and monitoring of neurodegenerative conditions.}\label{Neu}
\end{figure}

Alzheimer’s disease (AD), Parkinson’s disease (PD), and multiple sclerosis (MS) are among the most prevalent neurodegenerative disorders, characterized by progressive loss of neurons and irreversible functional decline, as shown in Figure \ref{Neu}. The exact etiology of these diseases remains uncertain, although genetic, environmental, and metabolic factors are believed to contribute to their pathogenesis. Despite advances in research, early and accurate diagnosis remains a major challenge, as symptoms often manifest only after substantial neuronal damage has occurred. Current diagnostic methods rely primarily on clinical assessment, neuroimaging, and biomarker analysis, but these techniques often fail to detect early-stage neurodegeneration before significant cognitive, motor, or sensory impairment develops.

Given the anatomical and embryological continuity between the retina and the central nervous system (CNS), retinal imaging has emerged as a non-invasive and promising tool for detecting early neurodegenerative changes. The retina provides a direct high-resolution window for studying neuronal and vascular changes associated with neurodegenerative diseases, offering a unique opportunity to assess structural and functional deterioration before clinical symptoms become apparent. Various retinal biomarkers, including thinning of the retinal nerve fiber layer (RNFL), loss of ganglion cells, microvascular alterations, and amyloid deposits, have been associated with neurodegenerative processes, making retinal imaging a valuable adjunct in early disease detection.

Technological advances, including optical coherence tomography (OCT) and optical computed tomography (OCTA), have revolutionized retinal imaging, allowing for precise visualization of retinal microstructures and vasculature. OCT enables the quantitative assessment of RNFL thickness, which is often reduced in AD, PD, and MS due to progressive neuronal degeneration. OCTA further improves this evaluation by detecting microvascular abnormalities that reflect cerebral perfusion deficits commonly observed in neurodegenerative diseases. In addition, emerging applications of machine learning (ML) algorithms have significantly improved automated retinal analysis, providing more accurate, efficient, and scalable diagnostic tools.

This section explores the pathological changes in the retina associated with neurodegenerative diseases and discusses the potential of retinal imaging as an early diagnostic biomarker. Using non-invasive imaging technologies and AI-driven analytical models, retinal assessment could bridge the gap in preclinical neurodegenerative disease detection, allowing earlier intervention and improved patient outcomes.

\subsection{Alzheimer’s Disease (AD) and the Retinal Biomarker}

AD is the most prevalent form of dementia, marked by progressive cognitive decline, memory impairment, and widespread neuronal degeneration. The disease is driven by the accumulation of amyloid beta plaques and neurofibrillary tangles, leading to synaptic dysfunction and loss of brain volume. One of the key challenges in the management of AD is that diagnosis is typically made only after noticeable cognitive symptoms emerge, by which time significant and irreversible neuronal damage has already occurred. This late stage identification underscores the need for early detection strategies that can identify preclinical markers before substantial neurodegeneration occurs.

Given its embryological and functional connection to the CNS, the retina has become a promising avenue for the early detection of AD-related changes. The structural and microvascular similarities between the retina and the cerebral vasculature make it a potential site for assessing neurodegenerative processes. Since the retina is easily accessible for imaging, researchers have explored retinal biomarkers as early indicators of AD pathology. Advancements in non-invasive imaging modalities, such as optical coherence tomography (OCT) and retinal fundus photography, have facilitated the detection of subtle retinal alterations linked to AD.

Numerous studies have reported retinal nerve fiber layer (RNFL) thinning, ganglion cell layer atrophy, and microvascular abnormalities in individuals with early-stage AD. These retinal changes mirror the neuronal degeneration observed in the brain, reinforcing the hypothesis that retinal structural deterioration reflects cerebral neurodegeneration. In addition, OCT and OCTA have identified reduced retinal blood flow and vascular density, indicating that microvascular dysfunction in the retina may serve as an early marker of AD progression.

A particularly significant retinal biomarker for AD is the presence of amyloid-beta deposits in retinal tissues, a pathology that mirrors the hallmark characteristics of AD in the brain. Recent studies using fluorescence imaging and hyperspectral retinal scanning have demonstrated that amyloid-beta accumulates in the retina, suggesting that retinal imaging could serve as an innovative diagnostic tool for AD. Detecting these deposits non-invasively provides a promising approach to monitoring disease progression and identifying high-risk individuals.

Incorporation of AI and machine learning (ML) in retinal imaging has further improved the accuracy and efficiency of AD detection. AI-based models can analyze the thickness of the retinal layer, vascular alterations, and amyloid beta deposits, offering a more objective and scalable diagnostic solution. These advances could significantly improve early AD screening, allowing risk assessment before cognitive decline becomes apparent.

Given the increasing prevalence of AD and the limitations of current clinical diagnostic methods, retinal biomarkers present a non-invasive, cost-effective, and widely accessible alternative for early disease detection and monitoring. Although more research is required to validate the use of retinal imaging in clinical practice, its potential to identify preclinical neurodegenerative changes offers a transformative opportunity in the early diagnosis and management of AD.

\subsubsection{Retinal Structural Changes in Alzheimer’s Disease}

The retina has emerged as a significant focus in understanding the pathogenesis of AD, as it provides a non-invasive, accessible window into neurodegenerative processes. Since the retina and the brain share common embryological origins, neuronal degeneration in the central nervous system (CNS) is often mirrored by structural alterations of the retina. Studies have shown that individuals with AD and AMD exhibit characteristic retinal changes, one of the most notable being the thinning of RNFL. This reduction in RNFL thickness can be measured precisely using OCT, making it a valuable tool for detecting early-stage neurodegeneration \cite{Berisha2007}.

Retinal degeneration in AD has been linked to synaptic loss and neuronal death, particularly in areas such as the hippocampus, which plays a crucial role in memory formation and cognitive function. As the disease progresses, the thinning of RNFL and ganglion cell atrophy become increasingly evident, corresponding to cortical neurodegeneration. These findings suggest that retinal changes could serve as early indicators of AD, which could offer a means of identifying neurodegeneration before cognitive symptoms become apparent.

A comprehensive review by Coppola \emph{et al.}\cite{Cipolla2024} confirmed that AD patients consistently exhibit significantly lower RNFL thickness compared to healthy individuals \cite{Cipolla2024}. In particular, the study found that the inferior quadrant of the retina is more affected than the superior quadrant, suggesting region-specific neurodegeneration patterns. This asymmetry may provide further insight into localized cortical atrophy in AD and could be explored as a biomarker for early diagnosis.

The potential of OCT in the detection of AD lies in its ability to provide high-resolution quantitative measurements of retinal structures, enabling an objective assessment of neurodegenerative changes. Unlike invasive diagnostic techniques such as CSF analysis or PET imaging, OCT offers a rapid, cost-effective, and non-invasive alternative to evaluate early AD-related retinal changes. As research progresses, the integration of OCT findings into routine clinical evaluations could significantly improve the early detection of AD and risk stratification.

The emergence of AI and ML in the analysis of retinal imaging has further expanded the diagnostic potential of OCT. AI-powered systems can detect subtle retinal changes, vascular abnormalities, and neurodegenerative markers with greater precision, facilitating the early identification of individuals at risk for AD. The implementation of automated AI-based retinal screening tools could streamline early detection strategies, allowing more effective intervention before substantial cognitive decline occurs.

Given the correlation between RNFL thinning, ganglion cell loss, and hippocampal atrophy, retinal imaging holds great promise as an early biomarker of AD. The integration of OCT and AI-driven analysis could revolutionize early AD detection and monitoring, providing a non-invasive means of tracking disease progression. Future research should aim to validate retinal biomarkers through large-scale longitudinal studies, ensuring that retinal imaging becomes a standard diagnostic tool for the assessment of neurodegenerative diseases.

\subsubsection{Retinal Amyloid Plaques as a Biomarker for Alzheimer’s Disease}

Research on retinal biomarkers for AD has revealed that beta-amyloid ($(A_\beta)$) deposits in retinal tissue may serve as early indicators of neurodegeneration. $(A_\beta)$ accumulation in the brain is a well-established pathological hallmark of AD, and recent findings suggest that similar amyloid plaques form in the retina long before cognitive symptoms appear. These retinal deposits offer a promising non-invasive approach for identifying individuals at risk for AD, potentially enabling earlier diagnosis and intervention before significant neurodegeneration occurs.

A foundational study by Koronyo-Hamaoui \emph{et al.} \cite{Koronyo2017} identified $(A_\beta)$ deposits in the retinas of AD patients, demonstrating a strong correlation between retinal amyloid accumulation and postmortem brain pathology \cite{Koronyo2011}. These findings reinforced the concept of the retina as an accessible extension of the CNS, where disease-related changes can be detected in a non-invasive manner. This discovery has led to further exploration of retinal imaging techniques to visualize amyloid deposits in living individuals, making retinal evaluation a potential diagnostic tool for preclinical detection of AD.

Advances in retinal imaging technologies have significantly improved the ability to detect amyloid plaques in vivo. Hamzah \emph{et al.} \cite{Hamzah2024} demonstrated that a combination of OCT and fluorescence imaging could successfully identify retinal amyloid plaques in patients with AD, confirming the feasibility of non-invasive amyloid detection \cite{Hamzah2024}. Their study also indicated that amyloid deposits were visible in individuals with mild cognitive impairment (MCI), a known precursor to AD, highlighting the potential of retinal screening for the identification of early-stage diseases.

The appeal of retinal amyloid imaging lies in its non-invasiveness, accessibility, and cost-effectiveness compared to traditional methods such as CSF analysis and PET imaging. Although PET scans remain the gold standard for detecting the cerebral amyloid burden, they are expensive, require specialized facilities, and expose patients to radiation. In contrast, retinal imaging offers a rapid, non-invasive, and potentially scalable alternative, allowing for a wider population screening and early risk assessment.

The integration of AI and ML into OCT-based amyloid detection has further improved diagnostic accuracy and efficiency. AI algorithms can process high-resolution retinal scans, identifying subtle amyloid deposits with greater precision and reliability. The use of automated image analysis could improve the sensitivity and specificity of AD detection, allowing earlier diagnosis and more effective disease monitoring.

Given the increasing prevalence of AD worldwide, retinal amyloid imaging represents a groundbreaking advance in early disease detection. Ongoing research should focus on validating these findings through longitudinal studies, improving retinal imaging protocols, and developing standardized diagnostic criteria. If successfully integrated into clinical practice, the detection of retinal amyloid plaque could become a key component of early diagnosis of AD and risk stratification, ultimately contributing to more effective disease management and therapeutic interventions.

\subsection{Parkinson’s Disease (PD) and Its Effect on Vision}

PD is the second most prevalent neurodegenerative disorder, mainly associated with the degeneration of dopaminergic neurons in the substantia nigra, a critical region of the brain responsible for the regulation of motor control. Progressive loss of these neurons leads to characteristic motor symptoms, including bradykinesia, resting tremors, rigidity, and postural instability. Although PD has long been classified as a movement disorder, recent studies indicate that it affects multiple regions of the brain, leading to non-motor symptoms that affect various physiological functions. Among these, visual impairments have become a significant but often overlooked aspect of PD pathology.

Research suggests that ocular dysfunction in PD is associated with dopamine depletion in the retina, which plays an essential role in contrast sensitivity, motion perception, and light adaptation. Since the retina contains dopaminergic neurons, progressive loss of dopamine-producing cells contributes to visual disturbances, including reduced contrast sensitivity, impaired color vision, visual hallucinations, and difficulty adjusting to low-light environments. These deficits can significantly affect daily activities, increasing the risk of falls, accidents, and mobility difficulties in people with PD.

The impact of PD on retinal function has drawn attention to the potential of retinal imaging as a non-invasive method for detecting and monitoring neurodegeneration. Studies using optical coherence tomography (OCT) and OCT angiography (OCTA) have demonstrated thinning of the retinal nerve fiber layer (RNFL) and the ganglion cell layer (GCL) in patients with PD, indicating that retinal changes may reflect the neurodegenerative processes that occur in the brain. The reduction in retinal thickness observed in patients with PD is particularly notable in the inner retinal layers, suggesting that retinal imaging could provide valuable information on PD progression.

Furthermore, research has highlighted microvascular abnormalities in the retina of patients with PD, supporting the hypothesis that vascular dysfunction plays a role in the pathology of the disease. The altered blood flow and microcirculation in the retinal capillaries suggest possible disruptions in cerebral perfusion, strengthening the link between vascular health and neurodegeneration. Given these findings, retinal biomarkers can serve as an early indicator of PD, allowing early diagnosis and monitoring of disease progression.

The integration of artificial intelligence (AI) and machine learning (ML) in retinal imaging has further advanced the ability to detect subtle retinal changes associated with PD. AI-powered analysis of OCT scans can provide automated detection of structural abnormalities, improve diagnostic accuracy, and facilitate early intervention. These technologies promise to improve the assessment of PD and track disease progression over time.

As research into PD-related vision impairment continues to evolve, retinal imaging presents an opportunity for early, non-invasive detection of neurodegenerative changes. More studies are needed to establish standardized retinal biomarkers and validate their clinical application in the diagnosis and monitoring of PD. With ongoing advances in OCT-based imaging and AI-driven analysis, retinal assessment could become an essential tool in the early detection of PD, disease monitoring, and personalized treatment approaches.

\subsubsection{Macular Retinal Layer Thinning as a Biomarker in Parkinson’s Disease}

The thinning of the retinal layer in the macular region has emerged as a valuable biomarker for assessing the severity and progression of Parkinson's disease (PD). The macula, which contains a dense concentration of retinal ganglion cells (RGCs) and photoreceptors, is particularly susceptible to neurodegenerative changes in PD. Several studies have shown that patients with PD experience significant thinning of the retinal nerve fiber layer (RNFL) and ganglion cell layer (GCL) in the macular area, correlating with the duration and severity of the disease. These findings suggest that retinal imaging could provide an objective tool for monitoring PD progression and assessing neurodegenerative damage over time.

A theoretical model proposed by Sabanayagam \cite{Sabanayagam2020} suggests that nearly all individuals with PD exhibit telangiectasia in the mega-reflex, indicating that structural retinal abnormalities may be a common characteristic of the disease. This supports the idea that retinal imaging could provide information on multiple aspects of PD pathology, making it a promising approach for both diagnosis and disease tracking.

Further supporting these findings, Güngör \emph{et al.}\cite{Guymer2019}  conducted an in-depth study revealing that patients with advanced PD showed a pronounced thinning of the inner nuclear layer and progressive loss of RNFL, and these structural changes aligned with the Unified Parkinson Disease Rating Scale (UPDRS) scores \cite{Guymer2019}. This correlation indicates that quantifying retinal layer thinning through imaging techniques could serve as a valuable biomarker for disease staging and provide a non-invasive means of tracking neurodegeneration.

The use of optical coherence tomography (OCT) and OCT angiography (OCTA) has significantly improved the ability to detect and quantify retinal layer thinning, offering a precise and non-invasive method for assessing neurodegeneration in PD patients. These technologies allow for detailed visualization of structural changes in the macular region, enabling early detection and facilitating more accurate monitoring of disease progression.

The integration of artificial intelligence (AI) and machine learning (ML) in retinal imaging has further enhanced the potential for automated macular scan analysis, improving early detection and tracking of thinning of the retinal layer. The application of AI-driven retinal imaging could allow real-time monitoring of disease progression, support personalized treatment approaches, and provide quantifiable biomarkers for clinical trials evaluating new PD therapies.

Future research should focus on validating retinal biomarkers in large-scale longitudinal studies, refining imaging techniques, and integrating retinal assessments into routine PD evaluations. If confirmed through further studies, macular retinal layer thinning could become an essential biomarker for early diagnosis, disease progression tracking, and therapeutic monitoring in patients with PD, ultimately improving patient outcomes and advancing treatment strategies.

\subsubsection{Retinal Blood Vessels in Relation to Parkinson’s Disease}

Although vascular alterations in the retina of patients with PD have not been studied as extensively as in AD, growing evidence suggests that retinal microvascular changes are also present in individuals with PD. The retina, being a highly vascularized structure, depends on a well-regulated neurovascular unit to maintain proper function. Disruptions in retinal blood flow and vessel integrity may indicate neurovascular dysfunction, which has been increasingly recognized as a contributing factor in PD pathology \cite{Kwapong2018}.

Recent findings from OCTA studies have shown a notable reduction in vessel density and altered retinal perfusion patterns, particularly in the macular region, among PD patients \cite{} \cite{Pillai2016}. These vascular alterations suggest an impairment in the retinal microcirculation, which could contribute to the visual deficits observed in PD. Neurovascular regulation dysfunction may reflect similar cerebrovascular abnormalities that occur in the brain of PD patients, further strengthening the connection between vascular health and neurodegenerative progression.

The observed vascular changes in the retina could be related to systemic endothelial dysfunction, a well-documented factor in the progression of PD. Chronic neuroinflammation and oxidative stress, both central mechanisms in PD pathology, have been proposed to contribute to vascular abnormalities in the retinal microcirculation. Reduced capillary density and altered blood flow dynamics may affect retinal ganglion cells (RGCs) and other neuronal structures, exacerbating visual disturbances in individuals with PD \cite{Matlach2018}.

Although the relationship between retinal vascular changes and PD is still under investigation, these findings highlight the potential role of retinal imaging in assessing disease progression. The application of OCT and OCTA in PD research could provide valuable insights into microvascular health, offering a non-invasive biomarker for the early detection and monitoring of PD-related neurodegeneration. Future studies should focus on establishing standardized retinal vascular biomarkers and exploring their clinical significance in disease progression and response to treatment \cite{Murueta2021}.

With advances in AI-driven image analysis, automated detection of vascular abnormalities in patients with PD may enhance diagnostic precision and allow real-time monitoring of retinal microvascular changes. Integration of retinal vascular assessments into routine PD evaluations could potentially aid in early diagnosis and personalized disease management, offering a new perspective on the neurovascular aspects of PD pathology.

\subsection{Retinal Omics in Multiple Sclerosis}
MS is a chronic autoimmune disorder that affects the CNS, leading to progressive demyelination, neurodegeneration, and axonal damage. Among its early clinical manifestations is ON, an inflammatory condition of the optic nerve that often results in visual disturbances and vision loss. Since ON is closely associated with MS, retinal imaging has gained significant importance in evaluating disease progression and evaluating treatment response. The ability to analyze retinal microstructure and vasculature non-invasively offers a unique opportunity to monitor neurodegenerative changes that occur in MS patients \cite{Wagner2020}.

The retina, which is a direct extension of the CNS, provides an accessible platform to detect early neurodegenerative changes related to MS pathology. With advances in retinal imaging and computational analysis, retinal omics, an approach that integrates imaging data with molecular and AI-driven analyses, is transforming how MS progression and treatment responses are studied. This technique allows for a detailed structural and functional assessment of the retina, making it a valuable tool for early detection and disease monitoring.

OCT and OCTA have been instrumental in uncovering retinal abnormalities in MS patients. Studies have shown that individuals with MS exhibit a reduction in RNFL and GCL, reflecting axonal and neuronal loss in the CNS. These changes are particularly prominent in patients with a history of ON, reinforcing the role of retinal imaging as a biomarker of MS-related neurodegeneration. In addition, vascular dysfunction in the retina, characterized by reduced capillary density and altered blood flow, has been identified in MS patients, suggesting a potential link between microvascular impairment and disease progression.

With the emergence of AI and machine learning (ML) algorithms, the precision of retinal imaging in MS diagnosis and monitoring has improved significantly. AI-powered analysis enables the detection of subtle changes in the retinal layer, allowing for more accurate disease staging and personalized treatment monitoring. The integration of AI-driven retinal analytics can further optimize early MS detection, enabling more proactive and targeted therapeutic interventions.

Future research should focus on standardizing retinal biomarkers for MS, refining retinal omics methodologies, and validating their clinical utility in longitudinal studies. If successfully integrated into routine clinical assessments, retinal imaging could become a cornerstone in MS treatment, offering a non-invasive, efficient, and cost-effective approach to tracking disease progression and treatment response.

\subsection{Fundus Imaging in Vascular Dementia}

Although fundus imaging has been widely used to identify retinal changes in AD, its application in other neurodegenerative diseases has been limited. However, recent research suggests that fundus imaging can provide valuable information on vascular dementia, a condition caused by damage to small blood vessels in the brain, leading to cognitive decline and neurological impairment \cite{Li2024}. Given that retinal microvasculature mirrors cerebral circulation, studying retinal vascular alterations could offer a non-invasive method for differentiating vascular dementia from other forms of dementia.

Vascular dementia is associated with microvascular damage, leading to characteristic retinal vascular abnormalities such as widened venules and narrowed arterioles. These changes reflect underlying cerebrovascular pathology, reinforcing the idea that retinal imaging could serve as a surrogate marker of vascular health in the brain \cite{Baba2021}. Since vascular dysfunction plays an important role in neurodegeneration, the analysis of the morphology of the retinal vessels may aid in early detection and disease classification.

A study by Kang \emph{et al.} (2021) investigated retinal vascular differences between patients with AD and vascular dementia, using fundus photography to compare vascular patterns \cite{Kang2020}. The findings revealed that AD patients exhibited narrower arterioles and less tortuous venules, whereas individuals with vascular dementia showed more pronounced vascular abnormalities. These results suggest that fundus imaging may help differentiate between various types of dementia, offering a cost-effective and non-invasive diagnostic approach.

Advancements in AI and ML-assisted retinal analysis further enhance the potential of fundus imaging in dementia classification and disease monitoring. Automated analysis of retinal vascular parameters could allow for the early detection of microvascular impairment, supporting timely intervention strategies. As research progresses, the integration of fundus imaging into dementia diagnostics could improve risk assessment, disease staging, and treatment planning.

Future studies should focus on validating retinal vascular biomarkers for vascular dementia, refining imaging techniques, and standardizing diagnostic criteria. If successfully incorporated into clinical practice, fundus imaging could become a key tool in dementia research, enabling a more accurate and early differentiation of neurodegenerative diseases.

\section{Advancements in Retinal Imaging and AI for Neurodegenerative Disease Detection}

The combination of advanced retinal imaging techniques with machine learning algorithms (ML) has significantly improved the ability to detect and monitor neurodegenerative diseases. Technologies such as optical coherence tomography (OCT), OCT angiography (OCTA), and fundus photography provide detailed images of retinal structures and vascular changes, which can be further analyzed using ML models. This automated approach enables the identification of subtle retinal abnormalities that may not be easily detected by human observers, allowing for early diagnosis and disease progression tracking in a non-invasive manner.

A study by Barriada \emph{et al.} (2022) demonstrated the potential of deep learning models in the analysis of retinal images for the detection of neurodegenerative diseases \cite{Barriada2022}. By incorporating OCT and fundus images from patients with different neurodegenerative conditions, the study achieved high diagnostic precision to predict disease status. Their findings indicated that retinal vascular patterns, particularly capillary plexus methylation, were strong predictors of Alzheimer's disease (AD) and Parkinson's disease (PD), reinforcing the role of retinal imaging as a biomarker of neurodegenerative disease progression.

AI-driven retinal analysis has also been instrumental in distinguishing neurodegenerative diseases based on retinal biomarkers. Automated systems have successfully identified disease-specific changes in retinal layer thickness, vascular integrity, and microvascular alterations, enabling precise classification of AD, PD, and multiple sclerosis (MS). The ability to track longitudinal changes in the structure and vasculature of the retina further allows the monitoring of disease progression and the evaluation of treatment responses.

The application of AI in retinal imaging is also improving the efficiency of large-scale screening programs for neurodegenerative diseases. Since retinal imaging is widely accessible and non-invasive, AI models trained on large datasets can process retinal images rapidly, improving early diagnosis rates while reducing the dependence on invasive procedures such as cerebrospinal fluid analysis and positron emission tomography (PET) scans.

Future research should focus on refining AI models to improve their accuracy in detecting early neurodegenerative markers and standardizing AI-assisted retinal diagnostics across different imaging platforms. If successfully integrated into clinical practice, AI-enhanced retinal imaging could transform neurodegenerative disease management, facilitating early intervention, monitoring disease progression, and personalized treatment planning.
\begin{longtable}{@{} p{2.5cm} p{7cm} p{2.5cm} p{2.5cm} @{}}
  \caption{Summary of findings on retinal biomarkers in neurodegenerative diseases}
  \label{tab:retinal_biomarkers_nd} \\
  \toprule
  \textbf{Reference} & \textbf{Key Findings} & \textbf{Disease} & \textbf{Clinical Significance} \\
  \midrule
  \endfirsthead

  \caption[]{Summary of findings on retinal biomarkers in neurodegenerative diseases (continued)}\\
  \toprule
  \textbf{Reference} & \textbf{Key Findings} & \textbf{Disease} & \textbf{Clinical Significance} \\
  \midrule
  \endhead

  \midrule \multicolumn{4}{@{}r}{\textit{Continued on next page}} \\ 
  \endfoot

  \bottomrule
  \endlastfoot

  MacGillivray \emph{et al.} \cite{MacGillivray2014}
    & Major neurodegenerative diseases (AD, PD, MS) are currently untreatable, highlighting the need for early detection tools.
    & AD, PD, MS
    & Emphasizes importance of early detection and monitoring. \\

  den Haan \emph{et al.} \cite{denHaan2017}
    & Systematic review and meta‐analysis showing significant thinning of RNFL and macular layers in AD and MCI vs.\ controls (SMD $\approx$ 1.0).
    & AD, MCI
    & Validates OCT‐derived retinal thinning as a non‐invasive biomarker for early AD. \\

  Cheung \emph{et al.} \cite{Cheung2022}
    & Deep‐learning on fundus photographs achieved AUROC=0.93 (93.2 \% sensitivity, 82.0 \% specificity) for Alzheimer’s detection.
    & AD
    & Enables scalable community screening for AD. \\

  Ma \emph{et al.} \cite{Ma2023}
    & OCTA revealed reduced macular vessel density and perfusion in AD and MCI; these vascular metrics correlated with cognitive scores (p $<$ 0.05).
    & AD, MCI
    & Supports OCTA‐derived vascular parameters as early biomarkers. \\

  Koronyo‐Hamaoui \emph{et al.} \cite{Koronyo2011}
    & Identified $\beta$-amyloid plaques in postmortem AD retinas and achieved noninvasive in vivo imaging of retinal plaques in AD model mice.
    & AD
    & Establishes retinal amyloid imaging for early diagnosis and monitoring. \\

  El‐Kattan \emph{et al.} \cite{ElKattan2022}
    & SD‐OCT showed thinning of RNFL, GCC, and macula in PD; structural deficits correlated with UPDRS motor scores and disease duration.
    & PD
    & Demonstrates structural biomarkers tracking PD severity. \\

  Elanwar \emph{et al.} \cite{Elanwar2023}
    & ff‐ERG and OCT revealed decreased bioelectrical activity and RNFL thinning in PD; changes correlated with UPDRS and duration (p $<$0.05).
    & PD
    & Highlights combined functional/structural biomarkers for PD monitoring. \\

  Doerr \emph{et al.} \cite{Doerr2011}
    & RNFL and macular thinning associated with lower brain parenchymal fraction on MRI in MS (r = 0.46–0.69, p $<$ 0.01).
    & MS
    & Positions OCT measures as non‐invasive CNS degeneration markers. \\

  Petzold \emph{et al.} \cite{Petzold2017}
    & Meta‐analysis demonstrating consistent RNFL and GCIPL thinning across MS subtypes; greater thinning predicted risk of disability worsening (HR $\approx$ 1.2 per 10 µm loss).
    & MS
    & Reinforces OCT layer segmentation as prognostic biomarker. \\

  Fisher \emph{et al.} \cite{Fisher2006}
    & RNFL thickness correlates with visual acuity, brain atrophy, and CNS degeneration in MS patients.
    & MS
    & Supports OCT as a monitoring tool for MS progression. \\
\end{longtable}

The findings presented in Table \ref{tab:retinal_biomarkers_nd} highlight the increasing importance of retinal biomarkers in the diagnosis and tracking of the progression of neurodegenerative diseases. The reviewed studies establish strong evidence that retinal imaging modalities, particularly OCT, OCTA, and fundus photography, provide essential insights into disease-specific alterations in retinal structures and vasculature. When integrated with AI-driven analysis, these imaging techniques allow for the early detection, classification, and continuous monitoring of conditions such as AD, PD, and MS.

    A key observation from the studies is that RNFL thinning is a consistent biomarker of neurodegeneration. Research by Fisher \emph{et al.} \cite{Fisher2006}, Doerr \emph{et al.} \cite{Doerr2011}, and Petzold \emph{et al.} \cite{Petzold2017} demonstrates a strong correlation between RNFL thinning and disease severity in MS and PD. The structural loss in RNFL and GCL aligns with cortical neurodegeneration, reinforcing the hypothesis that retinal imaging can serve as a surrogate marker for neurodegenerative progression. In the case of MS, the thinning of RNFL has been associated with brain atrophy and cognitive dysfunction, as highlighted by Elanwar \emph{et al.} \cite{Elanwar2023}, indicating its potential use to monitor the efficacy of treatment.

In addition, retinal vascular changes have been identified as early indicators of neurodegeneration. Studies suggest that small vessel alterations in the retina, detected through fundus photography and OCTA, can predict AD in the early stages. Furthermore, Jong \emph{et al.} (2021) demonstrated that retinal vascular abnormalities differ between AD and vascular dementia, with variations in venular and arteriolar structures, suggesting that retinal microcirculation analysis may help differentiate subtypes of dementia \cite{deJong2011}.

Retinal analysis powered by AI has also proven to be highly effective in disease classification and risk prediction. The study by Lee \emph{et al.} \cite{Lee2015} revealed that deep learning models applied to retinal images achieved high accuracy in predicting neurodegenerative diseases. Additionally, Petkos \emph{et al.} \cite{Pillai2016} demonstrated that AI analysis of retinal scans could predict the progression of MS, offering a novel approach to patient monitoring. These findings emphasize the potential for AI-assisted retinal evaluations to improve early diagnosis and individualized treatment plans.

One of the major advantages of retinal imaging is that it offers a non-invasive alternative to conventional diagnostic methods. Unlike CSF analysis or PET scans, which are costly and invasive, retinal imaging is widely accessible and can be seamlessly integrated into routine neurological evaluations. For example, research indicates that retinal imaging can detect beta-amyloid plaques, reinforcing its potential as a diagnostic tool for AD.

The clinical implications of these findings are substantial. Retinal imaging is emerging as a practical and scalable tool for the detection and monitoring of neurodegenerative diseases. Future research should focus on validating these biomarkers through longitudinal studies, refining AI-based diagnostic models, and standardizing retinal imaging protocols for widespread clinical use. If implemented successfully, retinal imaging could transform the management of neurodegenerative diseases by facilitating early intervention, continuous disease monitoring, and personalized treatment strategies.

\section{Retinal Imaging in Cardiovascular Disease Detection(CVD) and Risk Prediction}

CVDs encompass a broad spectrum of conditions, including ischemic heart disease, stroke, heart failure, and hypertensive disorders, all of which contribute significantly to global mortality. Although aging is an inevitable risk factor for cardiovascular deterioration, various internal and external stressors, such as genetic predisposition, lipid imbalances, glucose metabolism disorders, hypoxia, and hypertension, further accelerate the progression of the disease \cite{Flammer2013}. These conditions remain a leading cause of mortality worldwide, with a particularly high prevalence in the United States, as noted by Gupta \emph{et al.} \cite{Gao2024}. The increasing burden of CVDs requires the development of efficient and non-invasive diagnostic tools to assess cardiovascular risk and facilitate early intervention strategies.

\begin{figure}
  \centering
  \includegraphics[width=1\textwidth]{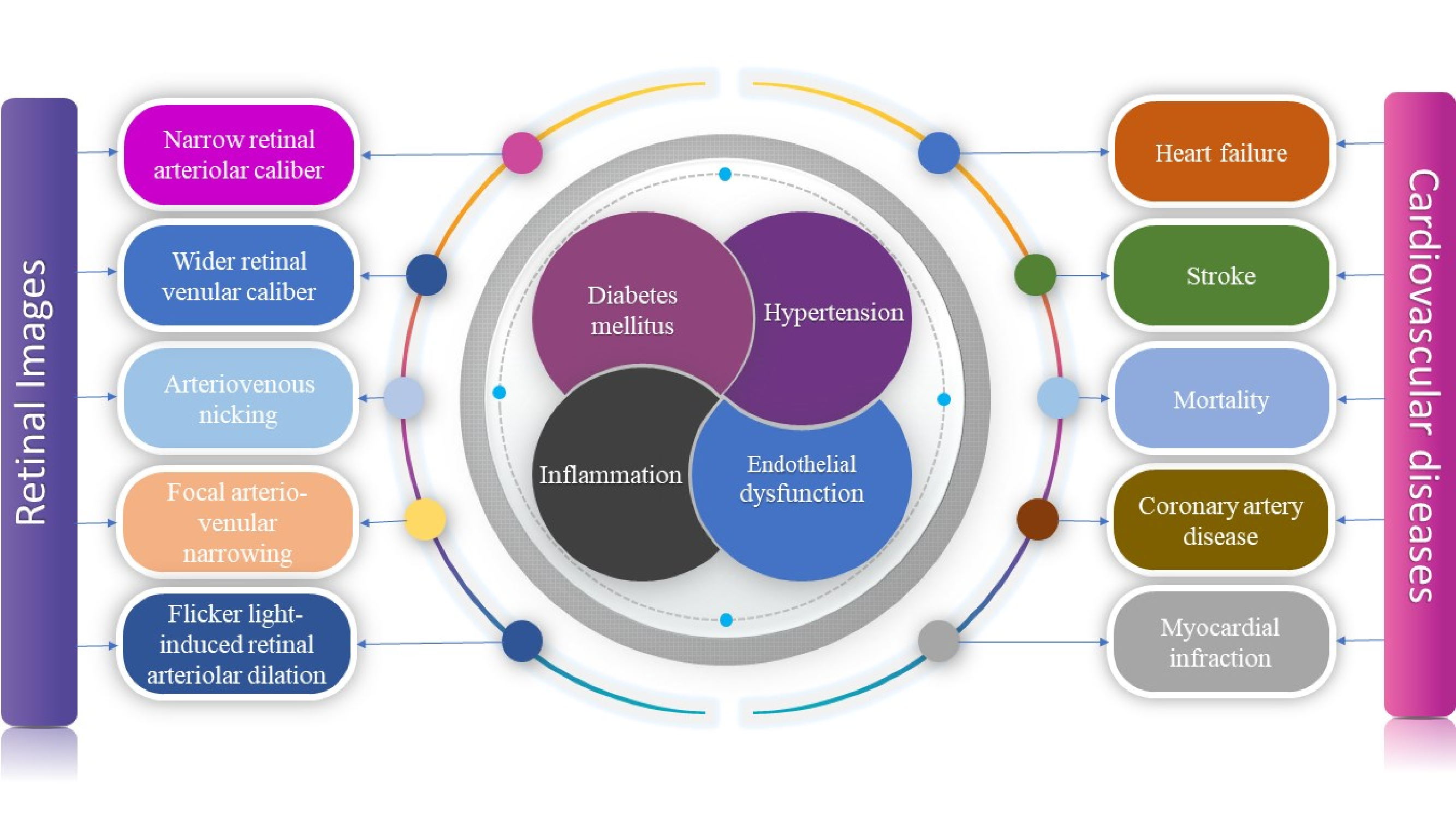}
  \caption{Key Retinal Features associated with heart diseases.}\label{heart}
\end{figure}

Recent advances in retinal imaging have allowed detailed visualization of microvascular alterations, reinforcing the notion that retinal microcirculation serves as a surrogate for systemic vascular health. The retinal vasculature shares anatomical and physiological similarities with the cerebral and coronary circulation, making it an essential site to detect early vascular changes that may precede clinical cardiovascular events. Figure \ref{heart} visually demonstrates the association between retinal vascular features and CVDs, emphasizing how specific retinal biomarkers correlate with systemic vascular dysfunction.

The left section of Figure \ref{heart} outlines key retinal microvascular abnormalities that are frequently observed in CVD individuals. These include narrowing of the retinal arterioles, widening of the venules, arteriovenous nicking, focal arteriovenular narrowing, and flicker-light-induced arteriolar dilation. These vascular changes are often indicative of hypertension, diabetes, and endothelial dysfunction, all of which contribute to cardiovascular morbidity. The central region of the figure presents these systemic risk factors, including hypertension, diabetes, inflammation, and endothelial dysfunction, as the main drivers of microvascular damage, influencing both the retinal and systemic vascular integrity.

The right section of Figure \ref{heart} connects retinal microvascular alterations to specific cardiovascular outcomes, including heart failure, stroke, coronary artery disease, myocardial infarction, and increased mortality. This connection highlights the clinical relevance of retinal imaging as a predictive tool for CVD risk assessment, as the same vascular processes that affect the retinal circulation often contribute to systemic cardiovascular pathology.

One of the key advantages of retinal imaging is its non-invasive nature, enabling direct visualization of vascular integrity and circulatory function. Techniques such as fundus photography, OCT, and OCTA facilitate the detection of subtle vascular abnormalities that may indicate underlying cardiovascular disease. For example, arteriolar narrowing and venular widening, as depicted in the left section of Figure \ref{heart}, are commonly associated with hypertension and endothelial dysfunction. Similarly, arteriovenous nicking and vessel tortuosity serve as early indicators of the progression of hypertensive and atherosclerotic disease. These findings suggest that retinal imaging could be incorporated into routine cardiovascular assessments, providing valuable information on an individual's vascular health and CVD susceptibility.

The integration of ML and AI has further expanded the potential of retinal imaging for the prediction of cardiovascular risk. AI-driven retinal analysis can automate the assessment of vascular patterns, detect microvascular abnormalities, and predict CVD risk factors with remarkable accuracy. Deep learning models have been developed to classify CVD risk based on retinal features, demonstrating superior performance compared to conventional risk prediction models. The ability to leverage large-scale retinal imaging datasets enables personalized cardiovascular risk assessment, improving early detection and targeted preventive strategies.

In addition, AI-assisted retinal imaging is being explored for automated identification of hypertensive and atherosclerotic changes. Studies have shown that AI models trained on retinal images can predict systemic health parameters, including blood pressure levels, cholesterol profiles, and myocardial infarction risk. These findings highlight the potential of integrating AI-powered retinal analysis into standard cardiovascular screening protocols, offering a scalable and efficient method for population-wide CVD risk assessment.

Despite these advances, more research is necessary to validate and standardize retinal biomarkers for the detection of cardiovascular disease. Large-scale longitudinal studies are required to establish the predictive value of retinal imaging in the stratification of CVD risk and the prediction of events. In addition, integrating retinal imaging into multidisciplinary healthcare frameworks will facilitate collaborative approaches to cardiovascular risk management and monitoring of disease progression.

As retinal imaging technology continues to evolve, its role in cardiovascular diagnostics is becoming increasingly prominent. By offering a non-invasive and accessible method for vascular health assessment, retinal imaging has the potential to revolutionize CVD risk prediction, early detection, and long-term disease monitoring. The convergence of advanced imaging techniques, AI-driven analytics, and cardiovascular research presents a transformative opportunity to improve patient outcomes and mitigate the growing global burden of cardiovascular diseases.

\subsection{Hypertension and Retinal Microvascular Alterations}

Hypertension is a widespread cardiovascular disease that has a profound impact on the retinal microvasculature, leading to structural and functional vascular changes. Since the retinal vasculature is an extension of the systemic circulatory system, hypertensive alterations in the small blood vessels of the retina provide a direct reflection of systemic vascular health. These changes can be effectively identified using non-invasive retinal imaging, offering valuable insight into early cardiovascular risk assessment and disease progression monitoring.

In individuals with hypertension, the retinal arterioles undergo progressive narrowing, leading to increased vascular resistance and reduced perfusion. Additionally, arteriovenous nicking, a hallmark of hypertensive retinopathy, occurs when stiffened arteries exert pressure on adjacent venules, resulting in venous compression and altered blood flow dynamics. Over time, microaneurysm formation can develop, indicating localized vascular stress and endothelial dysfunction. These retinal vascular changes not only serve as biomarkers of hypertensive damage, but also signal potential systemic vascular complications, including coronary artery disease, stroke, and heart failure.

Advancements in retinal imaging techniques, including fundus photography, OCT, and OCTA, allow a detailed evaluation of hypertension-induced microvascular changes. These modalities enable the detection of early-stage hypertensive retinopathy, even before clinical symptoms manifest, making them valuable tools for preventive cardiovascular care. The ability to visualize and quantify retinal vascular alterations provides a non-invasive means of monitoring hypertensive progression and evaluating treatment efficacy.

In addition, the integration of AI-driven analysis in retinal imaging has further enhanced the ability to detect and classify hypertensive vascular changes with high accuracy. Deep learning models trained on large datasets of retinal images can automatically assess vascular caliber, arteriovenous ratios, and microaneurysm distribution, allowing personalized risk stratification for people with hypertension-related vascular damage. The use of artificial intelligence in retinal analysis holds promise in improving the early detection and management of hypertensive complications, reducing the risk of serious cardiovascular events.

Despite these advances, more research is needed to standardize hypertensive retinal biomarkers and integrate retinal imaging into routine hypertension management protocols. Large-scale longitudinal studies will be essential to determine the predictive value of retinal vascular changes in relation to long-term cardiovascular outcomes. In addition, interdisciplinary collaboration between ophthalmology and cardiology could optimize the clinical application of retinal imaging to detect hypertension and monitor disease.

As retinal imaging technology continues to evolve, its role in the diagnosis and assessment of hypertension risk is becoming increasingly important. By offering a non-invasive, efficient and scalable method for evaluating microvascular health, retinal imaging has the potential to transform hypertension management, early intervention, and cardiovascular disease prevention strategies.

\subsubsection{Retinal Arteriolar Narrowing in Hypertensive Patients}

Retinal arteriolar narrowing is a well-documented vascular change observed in people with hypertension, serving as both a marker of existing cardiovascular stress and a predictor of future complications related to hypertension. Studies indicate that more than 90\% of hypertensive cases exhibit retinal arteriolar narrowing, highlighting its strong association with chronic systemic hypertension. This narrowing is mainly attributed to persistent elevation in blood pressure, which accelerates arteriosclerosis, leading to thickening of the vascular wall and luminal reduction. These structural changes increase vascular resistance, alter microcirculatory flow, and contribute to endorgan damage, particularly in the heart and brain.

Research has established that retinal arteriolar narrowing serves as an early predictor of hypertension, even in individuals who are not yet clinically diagnosed with the condition. The narrowing of these vessels indicates increased vascular stress, suggesting that retinal imaging may provide predictive information on the progression of hypertensive disease. In addition to hypertension, atherosclerosis and cerebrovascular disease have also been linked to the narrowing of the retinal arteriolar. A study by Liew \emph{et al.} \cite{Liew2008} found that patients with narrower retinal arterioles were at increased risk of ischemic heart disease, reinforcing the idea that the retinal vascular caliber could serve as a surrogate marker for atherosclerotic cardiovascular disease.

The relationship between retinal microvascular alterations and cardiac function has also been demonstrated in research by Cheung \emph{et al.}~\cite{Cheung2007}, which observed a negative correlation between retinal arteriolar narrowing and myocardial perfusion. This suggests that retinal vascular health is closely related to cardiac circulation, further supporting the role of retinal imaging in the evaluation of cardiovascular risk.

The advent of OCTA has significantly improved the ability to evaluate the retinal microcirculation non-invasively. This imaging modality provides high-resolution visualization of arteriolar narrowing and microvascular remodeling, offering a more comprehensive assessment of hypertensive retinal changes. Recent studies by Holwerda (2020) have highlighted the utility of OCTA in monitoring the long-term effects of hypertension on the retinal vasculature, allowing disease progression tracking and individualized risk assessment \cite{Holwerda2020}.

Using advanced retinal imaging techniques, clinicians can now detect early vascular changes associated with hypertension, enabling timely intervention and better management of cardiovascular risks. As further research continues to refine retinal biomarkers for diseases related to hypertension, retinal imaging can become an integral component of routine cardiovascular screening, helping in early detection, risk stratification, and personalized treatment strategies.

\subsubsection{Retinal Venular Dilation and Its Association With Stroke}

Although retinal arteriolar narrowing has been extensively associated with hypertension, retinal venular dilation has emerged as a significant predictor of cerebrovascular risk, particularly in relation to stroke and small vessel disease. According to studies, more than 30\% cases with increased venous width in macular region supply vessels have shown an elevated risk of ischemic and hemorrhagic stroke, suggesting a connection between chronic venous hypertension, vascular remodeling, and impaired cerebrovascular circulation \cite{Wong2001,Doubal2009}. These findings indicate that retinal venular dilation could serve as an early biomarker of cerebrovascular dysfunction, since it reflects increased venous pressure and endothelial stress, which may contribute to vascular damage to small vessels of the brain.

A study by Doubal \emph{et al.} (2009) found a strong association between retinal venular widening and lacunar stroke, a form of small vessel ischemic stroke caused by microvascular occlusion \cite{Doubal2009}. Vascular remodeling in the venules could indicate underlying venous pathology that affects both the retina and the cerebral circulation, where sustained vascular dilation and congestion affect microcirculatory function. Furthermore, structural changes in retinal venules near the foveal region suggest that venular dilation may be a manifestation of widespread endothelial dysfunction, potentially mirroring changes occurring in the brain microvasculature.

In addition to supporting this, McGeechan \emph{et al.} \cite{McGeechan2009} demonstrated that retinal venular dilation is a more significant predictor of stroke risk than arteriolar narrowing, confirming the importance of retinal imaging in assessing the risk of cerebrovascular disease. Venule widening has been associated with chronic inflammatory responses, endothelial injury, and increased blood-brain barrier permeability, all of which contribute to the progression of cerebrovascular diseases, including stroke.

Furthermore, research has indicated that retinal venular dilation is associated not only with stroke but also with cognitive decline and dementia, implying that retinal vascular abnormalities may reflect a wider microvascular dysfunction in the brain \cite{Witt2006}. This highlights the need for comprehensive retinal vascular assessments that incorporate venular changes as part of the cerebrovascular risk assessment. The potential of retinal imaging as a non-invasive biomarker of stroke and neurovascular diseases underscores its clinical utility in early detection, prevention, and stratification of the risk of cerebrovascular conditions.

\subsection{Atherosclerosis and Retinal Changes}
Atherosclerosis, a condition marked by plaque build-up within arterial walls, plays a crucial role in the onset of CVD such as CAD, stroke, and peripheral artery disease. This progressive disease leads to vascular stiffening, reduced blood flow, and an increased risk of arterial rupture, making it a major contributor to systemic vascular complications. Given that the retina provides direct visualization of the body’s microvasculature, retinal imaging has emerged as a valuable tool to detect early signs of atherosclerotic vascular changes, allowing pre-emptive intervention before severe cardiovascular events occur.

The retinal microvasculature mirrors systemic vascular health, offering an opportunity to observe early vascular alterations associated with atherosclerosis through non-invasive imaging techniques. Advanced imaging modalities such as fundus photography, OCT, and OCTA enable the detection of arteriolar narrowing, venular widening, increased vascular tortuosity, and endothelial dysfunction, all of which are indicative of subclinical atherosclerosis. These structural changes in the retinal circulation often precede clinically significant atherosclerotic complications, strengthening the role of retinal imaging in the evaluation of cardiovascular risk.

Studies have shown that narrowing of the retinal arterioles and an increase in the arteriovenous ratio are correlated with systemic atherosclerosis. These changes are primarily driven by chronic endothelial dysfunction, inflammation, and impaired vascular elasticity, which contribute to vascular rigidity and diminished autoregulatory capacity. The ability to quantify these microvascular changes in real time through retinal imaging provides a promising approach to early detection, allowing timely intervention and risk management.

Recent advances in AI and ML algorithms have further enhanced the ability to detect atherosclerosis-related retinal abnormalities with greater precision. AI-driven retinal image analysis enables automated detection of vascular alterations, improving the accuracy and efficiency of risk prediction models. Studies have shown that AI-powered retinal imaging can predict systemic vascular conditions, including coronary artery calcification and stroke risk, highlighting its potential in preventive cardiovascular healthcare.

As evidence continues to support the strong association between retinal vascular alterations and systemic atherosclerosis, incorporation of retinal imaging into routine cardiovascular evaluations could significantly improve early disease detection and risk stratification. Future research should focus on standardizing retinal biomarkers for atherosclerosis, optimizing AI-driven retinal diagnostics, and integrating these technologies into clinical practice for widespread screening and early intervention. With further advancements, retinal imaging is poised to become a critical tool in the early identification and management of atherosclerosis, ultimately improving cardiovascular outcomes and reducing mortality rates.

\subsubsection{Retinal Atherosclerosis and Coronary Artery Disease}
The association between retinal atherosclerosis and CAD has been well established, with studies indicating that retinal arteriolar narrowing, thickening of the vascular wall, and plaque formation in the retina are key markers of systemic atherosclerosis. Microvascular changes in the retina reflect the underlying coronary pathology, highlighting the potential of retinal imaging as a diagnostic tool for the detection of CAD. A study by Wong \emph{et al.} \cite{Wong2001} examined patients in the CAC Austin cohort, revealing a significant correlation between coronary artery calcification (CAC) and retinal vascular alterations, which were observed through fundus imaging. These findings reinforced the idea that retinal vascular degeneration is related to metabolic and inflammatory changes that occur in the larger coronary arteries, providing a window into systemic vascular health.

The study also confirmed that ocular microvascular abnormalities reflect oxidative stress (OS), making them a valuable addition to the risk assessment of CAD. Australian research has expanded on these relationships, demonstrating that chronic retinal microvascular changes correspond to long-term systemic vascular damage and decreased perfusion to ocular structures \cite{Wong2001}. Furthermore, Wang \emph{et al.} \cite{Wagner2020}) provided evidence supporting a mechanistic connection between OS-driven nephron dysfunction and CAD-related stenosis, suggesting that retinal microcirculation alterations are indicative of systemic microvascular compromise.

The progression of CAD is known to be multifactorial and involves the formation of arterial plaques, inflammation, and endothelial dysfunction, all of which are reflected in the microvascular remodeling of the retina. The implementation of advanced imaging techniques has allowed for better characterization of retinal vascular changes, reinforcing their value in detecting systemic pathology related to CAD. Studies using multimodal imaging approaches have demonstrated that retinal imaging can provide critical insight into the systemic impact of CAD, potentially improving early detection and intervention strategies.

A meta-analysis conducted by Wang \emph{et al.} (2018) further investigated coronary microvascular dysfunction, demonstrating a correlation between retinal arteriolar narrowing and reduced myocardial blood flow, suggesting that retinal vasculature alterations serve as a marker of systemic microangiopathy \cite{Wagner2020, McGeechan2009}. The link between retinal vascular abnormalities and atherosclerotic progression highlights the importance of retinal imaging in the assessment of the risk of cardiovascular disease.

In addition, retinal venular dilation has been associated with brain disease in small vessels and an increased risk of stroke, demonstrating the importance of retinal microvascular pathology as a systemic indicator of vascular health \cite{Doubal2009, Wong2001}. These findings support the integration of retinal imaging into clinical screening protocols to identify individuals at elevated risk for cardiovascular and cerebrovascular diseases. Using retinal imaging as a non-invasive diagnostic tool, clinicians can enhance early detection and personalized intervention strategies for patients susceptible to CAD and related vascular complications.

\subsection{AI and Machine Learning for Cardiovascular Risk Prediction Using Retinal Imaging}
Advancements in AI and ML have significantly improved cardiovascular risk assessment through retinal imaging, allowing early detection and personalized risk stratification. By analyzing retinal fundus images, OCT, and OCTA scans, AI-driven models can detect subtle vascular changes indicative of systemic disease. These technologies improve diagnostic accuracy by identifying microvascular abnormalities associated with cardiovascular risk factors such as hypertension, dyslipidemia, and hyperglycemia, even before clinical symptoms appear. Automation of these processes enables large-scale screening, positioning retinal imaging as an invaluable tool in preventive medicine.

The integration of AI into retinal imaging has shown remarkable potential to predict cardiovascular risk. A 2018 study analyzed retinal images from more than 280,000 subjects, demonstrating that AI models could accurately predict cardiovascular risk factors such as age, gender, and blood pressure \cite{Poplin2018}. The model achieved a confidence level of 0.70 in predicting major adverse cardiovascular events, exceeding many conventional risk models in predictive accuracy. These findings highlight the transformative role of AI in identifying people at high risk of cardiovascular disease, allowing timely intervention and preventive strategies.

Further advances in AI-driven retinal imaging have extended beyond basic risk assessment to predicting more severe cardiovascular conditions. The research by Zhang \emph{et al.} (2020) employed a deep learning model to identify hypertension, dyslipidemia, and hyperglycemia using images of the retinal fundus. The sensitivity to predict hypertension was 68. 8\%, while other cardiovascular risk factors demonstrated predictive precision that exceeded 70\% \cite{Zhang2020}. These findings validate the capacity of AI to improve risk stratification, allowing more precise detection of cardiovascular diseases and improving early intervention strategies.

The predictive power of AI has also been leveraged in the assessment of cardiovascular mortality. A study by Chang \emph{et al.} (2020) developed an artificial intelligence model that used images of the retinal fundus and carotid artery sonography to predict mortality related to atherosclerosis. The study revealed a strong correlation between retinal vascular abnormalities and cardiovascular death, which reinforces the need to incorporate AI-driven retinal imaging into standard clinical practice \cite{Chang2020}. These developments highlight the potential for integrating AI in retinal analysis, complementing conventional cardiovascular risk assessment tools, and improving patient outcomes through early disease detection.

The use of deep learning models to predict subclinical atherosclerosis and its progression has been further explored by Lieb \emph{et al.}\cite{Lieb2013}. Their study analyzed retinal images from diabetic patients, successfully classifying individuals into different atherosclerotic risk categories. This approach demonstrated that retinal imaging could predict the risk of atherosclerosis up to five years in advance, providing a reliable method to evaluate long-term cardiovascular risk \cite{Lieb2013}. AI-based retinal imaging has emerged as a valuable asset in predicting cardiovascular disease progression, offering a non-invasive means of assessing systemic vascular health.

Retinal vein occlusion has also been extensively studied for its association with systemic cardiovascular risk factors, including hypercholesterolemia and hypertension. As a vascular disorder characterized by retinal vein blockage, retinal vein occlusion has been increasingly linked to systemic vascular disease. Hyperlipidemia has been identified as a significant risk factor for retinal vein occlusion, strengthening its role as a potential early marker of systemic atherosclerosis. A study \cite{Yuan2022} found that people with elevated cholesterol levels had a higher probability of developing retinal vein occlusion, suggesting a connection between lipid metabolism dysfunction and retinal vascular blockages \cite{Yuan2022}. Klein \emph{et al.}\cite{Klein2007}  further supported this association, reporting that patients with retinal vein occlusion exhibited significantly higher serum lipid levels than those without the condition \cite{Klein2007}. Although the effectiveness of statin therapy in preventing retinal vein occlusion remains inconclusive, emerging evidence suggests that lipid-lowering medications may reduce the risk of retinal vascular occlusions. However, more research is needed to establish the long-term benefits of statins and other lipid-modifying therapies to maintain retinal vascular health \cite{Meier2013}.

In addition to hyperlipidemia, hypertension is a well-recognized risk factor for retinal vein occlusion. Studies indicate that hypertensive individuals are more susceptible to retinal vein blockages, with the severity of the disease often correlating with the extent of vascular damage. Research by Stettler \emph{et al.}\cite{Seidelmann2016} emphasized that people with retinal vein occlusion frequently exhibited elevated blood pressure and other hypertensive markers, suggesting a strong connection between hypertension and retinal vascular disease. The link between retinal vein occlusion and cardiovascular risk extends beyond hypertension, with newer studies identifying associations between retinal vein occlusion and broader cardiovascular outcomes, including stroke and coronary artery disease. The risk of ischemic stroke is significantly higher among patients with retinal vein occlusion, particularly when combined with other cardiovascular risk factors such as hypertension and hyperlipidemia. These findings highlight the potential of retinal vein occlusion as an early marker of systemic vascular disease and reinforce the importance of comprehensive cardiovascular evaluation in affected patients.

\begin{longtable}{@{} P{3cm} P{5cm} P{3cm} P{5cm} @{}}
  \caption{Summary of findings related to cardiovascular diseases and retinal parameters}
  \label{tab:retinal_cvd} \\
  \toprule
  \textbf{Reference} & \textbf{Key Findings} & \textbf{Disease} & \textbf{Clinical Significance} \\
  \midrule
  \endfirsthead

  \caption[]{\textit{(continued)}}\\
  \toprule
  \textbf{Reference} & \textbf{Key Findings} & \textbf{Disease} & \textbf{Clinical Significance} \\
  \midrule
  \endhead

  \midrule \multicolumn{4}{r}{\textit{continued on next page}} \\ 
  \endfoot

  \bottomrule
  \endlastfoot

  Poplin \emph{et al.} \cite{Poplin2018}
    & AI-based deep-learning models accurately predicted cardiovascular risk factors and major adverse cardiovascular events (MACE) from retinal images.
    & Cardiovascular disease
    & Demonstrates predictive utility of deep learning in cardiovascular risk assessment. \\

  Zhang \emph{et al.} \cite{Zhang2020}
    & AI models predicted hypertension, dyslipidemia, and hyperglycemia from retinal fundus images with high accuracy.
    & Cardiovascular risk factors
    & Supports use of AI for early detection of systemic metabolic conditions. \\

  Chang \emph{et al.} \cite{Chang2020}
    & Retinal imaging–based AI models predicted cardiovascular mortality related to atherosclerosis.
    & Cardiovascular disease
    & Enables non-invasive screening for elevated mortality risk. \\

  Matei \emph{et al.} \cite{Matei2017}
    & Hyperlipidemia was significantly associated with increased risk of retinal vein occlusion.
    & Retinal vein occlusion
    & Highlights link between lipid disorders and retinal vascular occlusions. \\

  Stettler \emph{et al.} \cite{Stettler2009}
    & Hypertensive patients exhibited higher prevalence of retinal vein occlusion and elevated systemic vascular risk.
    & Hypertension
    & Indicates need for cardiovascular evaluation in patients with retinal vein occlusion. \\
\end{longtable}

AI-driven retinal imaging represents a paradigm shift in cardiovascular risk assessment, positioning retinal biomarkers as essential tools for preventive medicine and early disease detection. With continued advances in artificial intelligence (AI) and deep learning, retinal imaging is set to become a cornerstone of cardiovascular disease detection, improving patient outcomes, and reducing the burden of cardiovascular-related mortality.

\section{AI-Driven Retinal Imaging for Systemic Disease Prediction}

Advancements in AI and ML have significantly improved the ability of retinal imaging to predict systemic diseases beyond traditional ophthalmic disorders. By analyzing retinal fundus images, OCT, and OCTA scans, AI-driven models can detect microvascular and structural changes associated with hematologic, psychological, and metabolic conditions. These approaches provide a non-invasive and scalable solution for early disease detection, risk assessment, and personalized healthcare interventions.

\subsection{Hematological Parameters}\label{sec:hematological}
\begin{longtable}{@{} P{3cm} P{8cm} P{4cm} @{}}
  \caption{Summary of AI‐based hematological predictions using retinal imaging}
  \label{tab:hematology} \\ 
  \toprule
  \textbf{Study} & \textbf{Key Findings} & \textbf{Clinical Application} \\
  \midrule
  \endfirsthead

  \caption[]{\textit{(continued)}}\\
  \toprule
  \textbf{Study} & \textbf{Key Findings} & \textbf{Clinical Application} \\
  \midrule
  \endhead

  \midrule \multicolumn{3}{r}{\textit{continued on next page}} \\ 
  \endfoot

  \bottomrule
  \endlastfoot

  Mitani \emph{et al.} \cite{Mitani2020}
    & AI models achieved AUC$>$0.75 in predicting anemia from fundus photographs.
    & Early non‐invasive anemia detection. \\

  Rebouças \emph{et al.} \cite{Reboucas2023}
    & AI‐based classification of anemia attained AUC$>$0.77.
    & Supports AI‐assisted hematological screening. \\

  Zhang \emph{et al.} \cite{Zhang2023}
    & Fundus‐image‐based models yielded low \(R^2\) (0.06–0.57) for hemoglobin and RBC counts.
    & Indicates need for multimodal analysis in hematological prediction. \\
\end{longtable}

Various hematologic parameters, including anemia, RBC count, hemoglobin concentration, and hematocrit, were analyzed using retinal fundus imaging. Deep learning models based on artificial intelligence have shown remarkable success in classifying anemia and related conditions. Mitani \emph{et al.}\cite{Mitani2020} reported an AUC greater than 0.75 for the detection of anemia using images of the retinal fundus, while Rim \emph{et al.}\cite{Reboucas2023}  achieved AUC values greater than 0.77 for the classification of anemia. These findings validate the potential of AI in hematological evaluations.

Further insights from Mitani \emph{et al.} \cite{Mitani2020} revealed that AI models trained to focus on specific retinal structures, such as the optic disc and macula, failed when these regions were obscured. This suggests that these anatomical features play a crucial role in the diagnosis of anemia. Interestingly, even when retinal images were blurred or scrambled, AI models could still predict anemia-related complications, indicating that distinct retinal characteristics are linked to systemic hematologic health.

Despite these advances, predicting continuous hematological parameters, such as RBC count and hemoglobin concentration, remains a challenge. Zhang \emph{et al.} \cite{Zhang2023} attempted to predict these values using deep learning models, but the results showed relatively low ${\rm{ }}{{\rm{R}}^2}$ values ranging from 0.06 to 0.57. This limitation may arise from complex physiological interactions between systemic blood parameters and retinal alterations. Future studies should integrate multimodal clinical data, including patient demographics, lifestyle factors, and medical history, to improve predictive accuracy.

\subsection{Psychological Disorders} \label{sec:psychological}

The role of retinal imaging in understanding psychological disorders has gained attention because of the direct connection of the retina to the central nervous system. The structural and vascular abnormalities observed in the retina have been associated with conditions such as depression, anxiety, schizophrenia, and bipolar disorder.

Almonte \emph{et al.} \cite{Almonte2020} demonstrated that patients with depression exhibited larger venular diameters, which are believed to reflect systemic inflammation, a factor known to contribute to mood disorders. Similarly, Almonte \emph{et al.} \cite{Almonte2020} reported a significant loss of microvascular density in individuals with generalized anxiety disorder, potentially due to prolonged stress-induced endothelial damage. These findings suggest that retinal imaging could serve as a biomarker of vascular stress for chronic psychological conditions.

Schizophrenia and bipolar disorder have also been associated with retinal vascular changes.  
Ascaso \emph{et al.} \cite{Ascaso2010} and Rodriguez-Jimenez \emph{et al.} \cite{Rodriguez2015} used OCT to examine the thickness of the RNFL in schizophrenia patients and found that a thinner RNFL was correlated with increased cognitive impairment and more severe psychotic symptoms.  
Appaji \emph{et al.} \cite{Appaji2019} reported that people with bipolar disorder exhibited a significantly increased retinal arteriolar tortuosity and reduced arteriolar diameters compared to healthy controls during mood episodes, supporting the role of microvascular dysfunction in affective disturbances.  
Meier \emph{et al.} \cite{Meier2013} identified an increase in venular tortuosity and a reduction in arteriolar caliber in patients with early-onset psychosis, suggesting that retinal microvascular alterations may serve as early biomarkers of psychotic disorders.

\subsection{Metabolic Diseases} \label{sec:metabolic}

Metabolic diseases, especially diabetes, are closely related to retinal abnormalities. Retinal imaging has long been used to detect diabetes-related complications such as DR, while also providing information on glycemic control through vascular changes.

DR remains a leading cause of blindness worldwide, making retinal imaging techniques such as OCT and fundus photography critical for its diagnosis and progression monitoring. These imaging modalities can capture key features of DR, including capillary leakage, occlusion, and neovascularization, facilitating early detection and treatment \cite{Almonte2020}. Furthermore, Sabanayagam \emph{et al.} \cite{sabanayagam2015retinal} demonstrated that retinal arteriolar narrowing and venular widening were associated with poor glycemic control, as indicated by elevated HbA1c levels. This suggests that retinal vascular changes could serve as biomarkers of metabolic dysregulation, alerting clinicians to early metabolic abnormalities.

Beyond DR, retinal imaging has been explored for predicting additional metabolic parameters, including fasting plasma glucose, insulin resistance, and diabetic peripheral neuropathy. Benson \emph{et al.} \cite{9175982} applied a deep learning model to predict diabetic peripheral neuropathy using images of the retinal fundus, achieving an accuracy of 0.89. This indicates that retinal imaging can go beyond the evaluation of retinopathy, encompassing broader metabolic evaluations. Similarly, Gerrits \emph{et al.} \cite{gerrits2020performance} used retinal imaging to predict testosterone levels in patients with metabolic disorders. Although the model effectively predicted gender, its accuracy in estimating testosterone was limited, suggesting that combining retinal imaging with other clinical data may improve predictive performance.

\begin{table}[htbp]
  \centering
  \caption{Summary of AI-based metabolic disease predictions using retinal imaging}
  \label{tab:metabolic}
  \begin{tabularx}{\textwidth}{@{} P{4cm} Y Y @{}}
    \toprule
    \textbf{Study} & \textbf{Key Findings} & \textbf{Clinical Application} \\
    \midrule
    Sabanayagam \emph{et al.} \cite{sabanayagam2015retinal}
      & Retinal vascular changes correlated with HbA1c levels.
      & Defines retinal biomarkers for glycemic control. \\
    Benson \emph{et al.} \cite{9175982}
      & AI predicted diabetic peripheral neuropathy with 0.89 accuracy.
      & Enables early non‐invasive neuropathy detection. \\
    Gerrits \emph{et al.} \cite{gerrits2020performance}
      & Retinal imaging estimated testosterone levels with limited accuracy.
      & Suggests need for multimodal analysis. \\
    \bottomrule
  \end{tabularx}
\end{table}

As AI-driven retinal imaging continues to evolve, its applications in systemic disease prediction expand beyond ophthalmology. Future research should focus on refining AI models, integrating multimodal clinical data, and validating findings through large-scale longitudinal studies. Integration of retinal imaging into routine medical practice could significantly improve early disease detection, risk stratification, and personalized treatment approaches, positioning retinal biomarkers as a cornerstone of precision medicine.

\subsection{Cross‐domain Challenges and Next Steps}
Several systemic domains in AI‐based retinal oculomics face distinct obstacles that must be addressed to advance clinical translation, summarized in Table~\ref{tab:gaps_challenges}. In the neurological domain, inconsistency in OCT/OCTA acquisition protocols and algorithmic bias threaten reproducibility and equitable performance between populations \cite{Munk2022,Mehrabi2019}. Cardiovascular applications are hampered by fragmented data-sharing infrastructures and a paucity of external validation studies, despite the availability of standards like SMART on FHIR for interoperability \cite{Mandel2016,Ting2019}. Metabolic disease screening suffers from limited access to high‐resolution imaging modalities in primary care settings and the absence of integrated multimodal datasets, although handheld fundus cameras show promise for point‐of‐care deployment \cite{Rego2024}. Renal oculomics has identified retinal biomarkers of chronic kidney disease but does not have longitudinal cohort studies to track predictive changes over time \cite{Sabanayagam2020,Berisha2007}. Hepatobiliary research is restricted by regulatory ambiguities and insufficient interdisciplinary collaboration, underscoring the need for ethical frameworks that align with clinical practice \cite{Char2018}. Finally, hematologic prediction models must overcome privacy concerns and dataset biases; Federated learning and expanded annotation efforts offer viable avenues forward \cite{Rieke2020,Mehrabi2019}. Addressing these cross‐domain challenges through standardized protocols, interoperable platforms, prospective study designs, and collaborative networks will be essential to realize the full potential of AI‐driven retinal oculomics.

\begin{table}[htbp]
  \centering
  \caption{Cross‐domain gaps and proposed solutions in AI‐based retinal oculomics}
  \label{tab:gaps_challenges}
  \begin{tabularx}{\textwidth}{@{} l X X @{}}
    \toprule
    \textbf{Systemic Domain} & \textbf{Key Challenges} & \textbf{Proposed Solutions} \\
    \midrule
    Neuro
      & Standardization of OCT/OCTA acquisition protocols; bias in AI training datasets
      & Develop international consensus guidelines; assemble multi‐ethnic neuro‐ocular imaging cohorts. \\

    Cardio
      & Data‐sharing limitations; lack of external validation across diverse populations
      & Establish interoperable repositories; conduct multicenter validation studies. \\

    Metabolic
      & Limited access to high‐resolution imaging in primary care; lack of multimodal data fusion
      & Deploy portable OCT devices; integrate retinal imaging with laboratory and clinical biomarkers. \\

    Renal
      & Absence of longitudinal tracking of retinal–renal changes
      & Initiate prospective cohort studies with serial retinal scans and renal function measures. \\

    Hepatobiliary
      & Limited interdisciplinary collaboration; regulatory hurdles for AI diagnostics
      & Form cross‐specialty consortia; align AI tools with FDA/CE regulatory frameworks. \\

    Hematology
      & Dataset bias and privacy concerns in sensitive hematological labeling
      & Implement federated learning approaches; expand diverse annotation efforts. \\
    \bottomrule
  \end{tabularx}
\end{table}

\section{Retinal Imaging in Renal and Hepatobiliary Disease Detection}

Chronic conditions such as \textbf{CKD} and hepatobiliary disorders are known to induce systemic microvascular injury, which can be detected in the retina. Retinal imaging is emerging as a non-invasive method for evaluating renal function and predicting liver disease progression. Microvascular alterations in the retina, such as venular dilation, arteriolar narrowing, and hemorrhages, have been associated with impaired kidney and liver function. This section discusses the role of retinal biomarkers in the detection of renal and hepatobiliary disorders and their potential to complement traditional diagnostic methods.

\subsection{Retinal Imaging for Chronic Kidney Disease Detection}

Retinal imaging has demonstrated significant potential in detecting \textbf{CKD} by assessing vascular abnormalities. Betzler \emph{et al.} \cite{Berisha2007} reported that \textbf{RFP} achieved an \textbf{AUC of 0.91} for internal validation, making it a strong predictor of CKD. However, external validation showed AUC values ranging from \textbf{ 0.73 to 0.84}, indicating the need for further validation in diverse populations. Additionally, Sabanayagam \emph{et al.}\cite{Sabanayagam2020} confirmed these findings in diabetic and hypertensive patients, suggesting that retinal imaging can serve as a viable screening tool for those at risk of progression of CKD.

Artificial intelligence-driven retinal analysis, when integrated with clinical parameters such as age, sex, and medical history, may improve CKD detection. This approach could reduce the dependency on invasive renal function tests and provide a cost-effective alternative for early screening.

\subsection{Retinal Vascular Changes Associated with Renal Dysfunction}

Renal impairment has been associated with significant vascular changes in the retina, including venular dilation and arteriolar narrowing. Kang \emph{et al.} \cite{Kang2020}  found that longitudinal retinal venular dilation was correlated with the decreasing estimated glomerular filtration rate (\textbf{eGFR}), serving as a potential early marker of kidney injury. Zhang \emph{et al.} \cite{Zhang2023}  further explored this association using deep learning algorithms, classifying CKD patients into low, medium and high risk categories based on retinal images.

These findings suggest that retinal vascular imaging could provide timely and non-invasive insight into renal dysfunction, helping early intervention and personalized disease management.

\subsection{Retinal Biomarkers for Hepatobiliary Disorders}

Hepatobiliary diseases, including cirrhosis, chronic hepatitis, liver cancer, and \textbf{NAFLD}, often remain undiagnosed until advanced stages. However, retinal imaging can detect early liver dysfunction through vascular abnormalities.

Xiao \emph{et al.} \cite{Xiao2017}  demonstrated that retinal imaging could predict hepatobiliary diseases with moderate accuracy, reporting AUC values between 0.62 (chronic viral hepatitis) and 0.84 (liver cancer). These findings suggest that retinal microvascular changes, such as vessel caliber alterations, may be influenced by systemic factors such as hypoalbuminemia, hyperammonemia, and portal hypertension.

%Additionally, Rim et al. \cite{}\cite{Rim2020} investigated the prediction of liver enzyme levels (AST and ALT) using RFP. Although their model produced an \textbf{R² value of 0.10}, indicating limited predictive power, it suggests that retinal features correlate with liver dysfunction, warranting further research.

\subsection{Retinal Vascular Complications in Cirrhosis and Portal Hypertension}

Portal hypertension, a secondary complication of cirrhosis, induces retinal vascular changes, including venular dilation and arteriovenous shunting. %Iwakiri et al\cite{Iwakiri2014} suggested that increased portal venous pressure adversely affects retinal microvasculature, making retinal imaging a potential biomarker for cirrhosis and hepatobiliary disorders.

These vascular alterations, combined with retinal hemorrhages and exudates, may indicate the body's compensatory response to hepatic dysfunction. If validated in clinical settings, retinal imaging could serve as an early detection tool for cirrhosis.

\begin{table}[htbp]
  \centering
  \caption{Summary of AI-based retinal imaging studies for renal and hepatobiliary diseases}
  \label{tab:renal_liver}
  \begin{tabularx}{\textwidth}{@{} P{4cm} Y Y @{}}
    \toprule
    \textbf{Study} & \textbf{Key Findings} & \textbf{Clinical Application} \\
    \midrule
    Betzler \emph{et al.} \cite{Berisha2007}
      & Retinal imaging predicted chronic kidney disease (CKD) with AUC up to 0.91.
      & AI-assisted CKD detection. \\
    Sabanayagam \emph{et al.} \cite{Sabanayagam2020}
      & Retinal imaging used as a screening tool for diabetic and hypertensive CKD patients.
      & CKD risk stratification. \\
    Kang \emph{et al.} \cite{Kang2020}
      & Retinal venular dilation linked to declining estimated glomerular filtration rate (eGFR).
      & Retinal markers for kidney injury prediction. \\
    Zhang \emph{et al.} \cite{Zhang2023}
      & AI classified CKD progression into low, medium, and high-risk groups.
      & AI-driven CKD monitoring. \\
    Xiao \emph{et al.} \cite{Xiao2017}
      & Retinal imaging predicted hepatobiliary diseases with AUC up to 0.84.
      & Liver disease screening tool. \\
    Rim \emph{et al.} \cite{Rim2020}
      & Retinal images correlated with AST and ALT enzyme levels.
      & Potential retinal biomarkers for liver dysfunction. \\
    Iwakiri \emph{et al.} \cite{Iwakiri2014}
      & Portal hypertension linked to retinal vascular abnormalities.
      & Retinal markers for cirrhosis and portal hypertension. \\
    \bottomrule
  \end{tabularx}
\end{table}

\subsection{Importance of Vessels Segmentaion}

Accurate segmentation of the retinal vasculature enables precise quantification of vessel caliber, tortuosity, and branching patterns, providing early biomarkers for systemic hypertension and cardiovascular risk \cite{khan2019boosting,khan2019generalized, khan2024esdmr}. High-sensitivity capillary segmentation facilitates the detection of microaneurysms and hemorrhages in diabetic retinopathy, supporting timely therapeutic intervention and monitoring of disease progression \cite{khan2019ggm,iqbal2022g, khan2021residual,khan2021rc}. Automated delineation of arteriolar and venular networks allows accurate measurements of the arteriovenous ratio, a key metric in hypertensive retinopathy and other microvascular disorders \cite{naqvi2019automatic,khan2020region}. Shallow vessel segmentation networks deliver efficient vessel maps under diverse imaging conditions, enhancing glaucoma detection through peripapillary vessel density analysis \cite{khan2020shallow,khan2020exploiting}. Hybrid deep architectures like TBConvL-Net integrate convolutional and transformer modules to enhance segmentation accuracy across heterogeneous imaging modalities \cite{iqbal2025tbconvl,khan2023feature}. The maturation of segmentation algorithms is essential for integrating AI-enhanced oculomics into precision medicine, allowing personalized disease prediction and monitoring \cite{khan2024esdmr,khan2021residual}.

\subsection{Conclusion and Future Directions}

Retinal imaging holds promise as a non-invasive diagnostic tool for detecting renal and hepatobiliary diseases. With AI and deep learning, retinal imaging can soon serve as an early marker, allowing proactive interventions and personalized treatment plans.

However, before retinal biomarkers can be fully integrated into clinical practice, several challenges must be addressed.
\begin{itemize}
    \item \textbf{External validation of AI models} in multiethnic and geographically diverse populations.
    \item \textbf{Integration of multimodal imaging} (for example, combining OCT, RFP, and AI-driven analysis) for enhanced diagnostic accuracy.
    \item \textbf{Further research on retinal vascular responses} to renal and liver dysfunctions, refining predictive models.
\end{itemize}

As retinal imaging technology advances, tools such as sweep source OCT and high-resolution fundus imaging can further improve disease detection capabilities, strengthening its role in systemic disease detection.

\section{Research Gaps}

Despite significant advances in AI-driven retinal imaging, several challenges must be addressed before it can be fully integrated into mainstream clinical practice. These challenges span multiple dimensions, including technical limitations, clinical applicability, and ethical concerns. Overcoming these barriers is essential for the successful deployment of AI-enhanced retinal imaging for the prediction of ocular and systemic diseases.

\subsubsection{Integration of Retinal Imaging into General Medicine}

A critical challenge in leveraging retinal imaging for systemic disease detection is the need for interdisciplinary collaboration. Currently, ophthalmology remains largely compartmentalized from other medical fields, despite the strong evidence linking retinal changes with conditions such as diabetes, cardiovascular disease, and neurodegenerative disorders. This separation has hindered the widespread adoption of retinal imaging as a standard diagnostic tool in general medicine.

To bridge this gap, greater integration between ophthalmologists, neurologists, cardiologists, nephrologists, and endocrinologists is necessary. Establishing standardized protocols for the incorporation of retinal imaging into routine systemic disease screening could enhance early diagnosis and patient management. In addition, cross-specialty collaboration would enable the development of multimodal AI models trained on comprehensive datasets that include retinal images, clinical parameters, and systemic biomarkers.

An additional challenge involves the development of centralized data-sharing frameworks. Currently, the lack of interoperability between healthcare systems prevents seamless access to retinal imaging data across different specialties. Creating shared databases with standardized imaging formats, integrated with electronic health records (EHR), could facilitate the training and validation of AI models. Ensuring data privacy and security in compliance with regulatory frameworks, such as the \textbf{General Data Protection Regulation (GDPR)} and the \textbf{Health Insurance Portability and Accountability Act (HIPAA)}, will be essential for the ethical implementation of such systems.

\subsubsection{Availability of High-Resolution Ophthalmic Imaging Datasets}

The availability of high-resolution retinal imaging datasets remains a significant limitation in the development of AI-based diagnostic models. Training deep learning algorithms requires large-scale, high-quality datasets that are diverse in terms of demographic representation, disease prevalence, and imaging modalities. However, acquiring such datasets is challenging due to several factors.

First, obtaining high-resolution retinal images requires specialized imaging equipment, such as \textbf{optical coherence tomography (OCT)}, \textbf{OCT angiography (OCTA)}, and high-resolution \textbf{fundus cameras}, which are not readily available in all healthcare settings. This lack of accessibility restricts the collection of large, standardized datasets for AI training.

Second, most existing datasets are geographically and ethnically biased, limiting the generalizability of AI models. Retinal vascular features and disease prevalence vary between populations, and models trained on homogeneous datasets may not perform well when applied to diverse groups of patients. Efforts to compile multiethnic, multicenter datasets are necessary to improve the robustness of AI-driven retinal analysis.

In addition, most current AI studies are based on retrospective datasets, which may not fully capture the progression of the disease over time. The development of longitudinal datasets, where retinal images are collected from the same individuals over extended periods, would provide valuable insights into disease dynamics and enable AI models to predict disease progression with higher accuracy.

\begin{table}[htbp]
  \centering
  \caption{Challenges in AI-based retinal imaging and potential solutions}
  \label{tab:research_gaps}
  \begin{tabularx}{\textwidth}{@{} P{5cm} Y @{}}
    \toprule
    \textbf{Challenge} & \textbf{Potential Solutions} \\
    \midrule
    Lack of interdisciplinary collaboration
      & Establish cross-specialty cooperation between ophthalmologists, neurologists, cardiologists, and nephrologists to integrate retinal imaging into systemic disease screening. \\
    Limited access to high-resolution imaging
      & Expand availability of advanced imaging technologies—such as OCT and fundus photography—in primary care and underserved regions. \\
    Data-sharing limitations
      & Develop standardized protocols and interoperable databases for integrating retinal imaging data with electronic health records. \\
    Bias in AI training datasets
      & Compile diverse, multi-ethnic datasets from international healthcare institutions to improve model generalizability. \\
    Lack of longitudinal data for disease progression prediction
      & Establish longitudinal studies to track retinal changes over time and enhance AI-based disease forecasting. \\
    Regulatory and ethical concerns
      & Implement robust data security measures and comply with privacy regulations (GDPR, HIPAA) to ensure ethical AI deployment. \\
    \bottomrule
  \end{tabularx}
\end{table}

\subsubsection{Standardization and Validation of AI Models}

A significant barrier to the widespread adoption of AI in retinal imaging is the lack of standardized validation protocols. Many AI models are trained on proprietary datasets with inconsistent labeling criteria, making it difficult to compare performance between different studies. The absence of external validation further raises concerns about model overfitting and clinical reliability.

To address this issue, the development of benchmarking datasets and standardized evaluation metrics is crucial. The implementation of multicenter clinical trials to assess the performance of the AI model in real-world settings will help establish confidence in AI-driven retinal analysis. Regulatory agencies, such as the \textbf{U.S. Food and Drug Administration (FDA)} and the \textbf{European Medicines Agency (EMA)}, should work with researchers to define criteria for the validation and clinical implementation of the AI model.

\subsubsection{Ethical Considerations and Patient Trust}

The deployment of artificial intelligence (AI) in retinal imaging raises several ethical concerns, particularly in terms of data privacy, algorithmic bias, and patient trust. AI models trained on biased datasets may produce biased predictions that disproportionately affect certain demographic groups. Additionally, there is an ongoing debate on the explainability of AI-driven decisions: many deep learning models function as "black boxes," making it difficult for clinicians to interpret their outputs.

To mitigate these challenges, there is a need for transparent AI systems that provide interpretable results. Explainable AI (XAI) techniques should be incorporated into retinal imaging models to offer clinicians insight into the decision-making process. In addition, patient education initiatives should be implemented to increase public awareness and acceptance of AI-driven diagnostics.

\subsection{Future Directions}

Although AI-powered retinal imaging has the potential to revolutionize disease detection and management, addressing existing research gaps is critical to its successful implementation. Future research should focus on:
\begin{itemize}
    \item Developing multi-center, ethnically diverse datasets to improve AI model generalizability.
    \item Integrating retinal imaging with systemic biomarkers to improve the accuracy of disease prediction.
    \item Standardization of AI validation protocols through regulatory collaboration.
    \item Advancing explainable AI methods to improve patient trust and transparency.
    \item Expanding access to retinal imaging in primary healthcare settings to enable early disease detection.
\end{itemize}

By tackling these challenges, retinal imaging can become a cornerstone of precision medicine, facilitating early detection and intervention for a wide range of systemic diseases.

\section{Conclusions}

Retinal imaging has become an invaluable tool for the early detection, monitoring, and stratification of the risk of systemic diseases, particularly those that affect the cardiovascular, neurological, and metabolic systems. The ability to non-invasively visualize retinal microvascular changes offers a unique opportunity to assess vascular and neural health in a way that is not possible with traditional diagnostic methods. Diseases such as hypertension, diabetes, and cardiovascular disorders exhibit characteristic retinal microvascular abnormalities that correlate with disease severity and progression. For example, in pregnant women with pro-atherosclerotic changes, early detection through retinal imaging could allow for timely intervention, potentially improving maternal and fetal health outcomes. Recent advances in imaging technologies such as adaptive optics (AO), optical coherence tomography (OCT), and fundus photography have greatly enhanced the resolution and efficiency of retinal imaging, allowing for precise visualization of vascular and neural structures.

The integration of artificial intelligence (AI) and machine learning (ML) into retinal imaging has further expanded its potential as a diagnostic and predictive tool. AI-driven models can analyze large datasets of retinal images with high precision, identifying subtle pathological changes that may not be evident using conventional diagnostic techniques. These computational methods enable automated screening, early disease detection, and improved risk stratification, thus playing a crucial role in personalized medicine. Studies have shown that AI-based retinal analysis can predict cardiovascular risk factors, neurodegenerative disease progression, and even hematological parameters such as anemia and hemoglobin concentration. This advancement underscores the potential of AI-powered retinal imaging to revolutionize precision medicine by providing early and accurate risk assessments for a wide range of diseases.

In addition to cardiovascular and metabolic disorders, retinal imaging has shown promise in the evaluation of neurodegenerative and psychiatric diseases. Changes in the retinal nerve fiber layer (RNFL), the ganglion cell layer (GCL), and the retinal microvasculature have been linked to conditions such as Alzheimer's disease, Parkinson’s disease, schizophrenia, and mood disorders. Given that the retina shares embryonic and physiological similarities with the central nervous system, its structural and vascular abnormalities can serve as potential biomarkers for neurodegenerative and cognitive disorders. Emerging research suggests that retinal imaging, particularly when combined with AI-based analysis, may play a crucial role in the early diagnosis and progression tracking of these conditions. However, more longitudinal studies are necessary to establish standardized biomarkers and validate their clinical significance.

Despite these promising advances, several challenges remain before retinal imaging can be fully integrated into routine clinical practice. Standardization of imaging protocols is crucial to ensure reproducibility and consistency in different healthcare settings. In addition, the development of robust analytical frameworks and validation studies is required to enhance the accuracy and reliability of retinal biomarkers. The successful clinical translation of AI-powered retinal analysis also depends on interdisciplinary collaboration among researchers, clinicians, and data scientists. Such collaborations will facilitate the refinement of AI models, optimize their predictive capabilities, and ensure their applicability to diverse populations. Moreover, ethical considerations regarding data privacy, algorithmic bias, and regulatory approvals must be addressed to promote widespread acceptance and trust in AI-driven retinal diagnostics.

Furthermore, while retinal imaging has demonstrated immense potential as a non-invasive screening tool, its role in disease prediction and management must be continuously refined through large-scale clinical trials. Integration of retinal imaging with other multimodal diagnostic approaches, such as genomic data and traditional clinical biomarkers, could further enhance its predictive value. Future research should focus on improving the resolution of the image, developing new biomarkers and refining AI algorithms to maximize the diagnostic and prognostic utility of retinal imaging. Additionally, the feasibility of incorporating retinal imaging into community health screening programs should be explored, particularly in resource-limited settings where access to specialized medical care is limited.

In summary, retinal imaging has the potential to transform modern healthcare by offering a non-invasive, cost-effective, and highly informative approach for detecting and monitoring a broad spectrum of diseases. The synergy between advanced imaging modalities, AI-driven analytics, and personalized medicine paves the way for earlier disease detection, better risk stratification, and improved patient outcomes. However, for retinal imaging to achieve widespread clinical adoption, ongoing research, technological advancements, and interdisciplinary collaboration will be essential. With further innovations and rigorous validation studies, retinal imaging could become a cornerstone of precision medicine, significantly contributing to global efforts in early disease detection, prevention, and management.

\end{document}